\documentclass[prd,eqsecnum,twocolumn,amsfonts,amssymb]{revtex4}

\usepackage{graphicx}

\usepackage{bm}

\newcommand{\beq}{\begin{equation}}
\newcommand{\eeq}{\end{equation}}
\newcommand{\beqs}{\begin{eqnarray}}
\newcommand{\eeqs}{\end{eqnarray}}
\newcommand{\lsim}{\mathrel{\raisebox{-
.6ex}{$\stackrel{\textstyle<}{\sim}$}}}

\begin{document}

\title{Physics of the Non-Abelian Coulomb Phase: 
Insights from Pad\'e Approximants}

\author{Thomas A. Ryttov$^a$ and Robert Shrock$^b$}

\affiliation{(a) \ CP$^3$-Origins, University of Southern Denmark, 
 Campusvej 55, Odense, Denmark}

\affiliation{(b) \ C. N. Yang Institute for Theoretical Physics and
Department of Physics and Astronomy, \\
Stony Brook University, Stony Brook, NY 11794, USA }

\begin{abstract}

  We consider a vectorial, asymptotically free SU($N_c$) gauge theory with
  $N_f$ fermions in a representation $R$ having an infrared (IR) fixed
  point. We calculate and analyze Pad\'e approximants to scheme-independent
  series expansions for physical quantities at this IR fixed point, including
  the anomalous dimension, $\gamma_{\bar\psi\psi,IR}$, to $O(\Delta_f^4)$, and
  the derivative of the beta function, $\beta'_{IR}$, to $O(\Delta_f^5)$, where
  $\Delta_f$ is an $N_f$-dependent expansion variable. We consider the
  fundamental, adjoint, and rank-2 symmetric tensor representations. The
  results are applied to obtain further estimates of $\gamma_{\bar\psi\psi,IR}$
  and $\beta'_{IR}$ for several SU($N_c$) groups and representations $R$, and
  comparisons are made with lattice measurements.  We apply our results to
  obtain new estimates of the extent of the respective non-Abelian Coulomb
  phases in several theories. For $R=F$, the limit $N_c \to \infty$ and $N_f
  \to \infty$ with $N_f/N_c$ fixed is considered. We assess the accuracy of the
  scheme-independent series expansion of $\gamma_{\bar\psi\psi,IR}$ in
  comparison with the exactly known expression in an ${\cal N}=1$
  supersymmetric gauge theory. It is shown that an expansion of
  $\gamma_{\bar\psi\psi,IR}$ to $O(\Delta_f^4)$ is quite accurate throughout
  the entire non-Abelian Coulomb phase of this supersymmetric theory.

\end{abstract}

\maketitle


\section{Introduction}
\label{intro_section}

The properties of a vectorial, asymptotically free gauge theory at an infrared
fixed point (IRFP) of the renormalization group (RG) in a conformally invariant
regime are of fundamental interest.  Of equal interest is the determination of
the infrared phase structure of such a theory. Owing to the asymptotic freedom,
one can perform reliable perturbative calculations in the deep ultraviolet (UV)
where the gauge coupling approaches zero, and then follow the
renormalization-group flow to the infrared.  This flow is described by the beta
function, $\beta = d\alpha/d\ln\mu$, where $\alpha=g^2/(4\pi)$, $g=g(\mu)$ is
the running gauge coupling, and $\mu$ is a Euclidean energy-momentum scale.
For a given gauge group $G$ and fermion representation $R$ of $G$, the
requirement of asymptotic freedom places an upper bound, denoted $N_u$, on the
number of fermions, $N_f$, transforming according to this representation.  The
UV to IR evolution of a theory is determined by $G$, $R$, and $N_f$. If $N_f$
is slightly less than $N_u$, then the RG flows from the UV to a weakly coupled
IRFP in a non-Abelian Coulomb phase (NACP, also called the conformal window) in
the infrared. The value of $\alpha$ at the IRFP is denoted $\alpha_{IR}$.  At
this IRFP, the theory is scale-invariant and is deduced to be conformally
invariant \cite{scalecon}.  Physical quantities at this IRFP can be expressed
perturbatively as series expansions in powers of $\alpha_{IR}$ (e.g.,
\cite{bvh,ps,bc,flir}).  However, beyond the lowest loop orders, the
coefficients in these expansions depend on the scheme used for regularization
and renormalization of the theory.  Consider an asymptotically free vectorial
gauge theory with gauge group $G$ and $N_f$ massless \cite{fm} Dirac fermions
$\psi_i$, $i=1,...,N_f$, in a representation $R$ of $G$, such that the RG flow
leads to an IRFP, and let $N_u$ denote the value of $N_f$ at which asymptotic
freedom is lost. Since $\alpha_{IR}$ becomes small as $N_f$ approaches $N_u$
from below, one can reexpress physical quantities as series expansions in the
manifestly scheme-independent variable
\beq
\Delta_f = N_u - N_f \ .
\label{deltaf}
\eeq
Some early work on this was reported in \cite{bz}-\cite{grunberg92}.  

Two quantities of considerable interest are anomalous dimensions of
(gauge-invariant) fermion bilinear operators and the derivative of the beta
function, evaluated at the IRFP,
\beq
\frac{d\beta}{d\alpha} \Big |_{\alpha=\alpha_{IR}} \equiv \beta'_{IR} \ . 
\label{betaprime_def}
\eeq
These are physical quantities and hence are scheme-independent
\cite{gross75}. The derivative $\beta'_{IR}$ is equivalent to the anomalous
dimension of ${\rm Tr}(F_{\mu\nu}F^{\mu\nu})$, where $F_{\mu\nu}^a$ is the
field-strength tensor of the theory \cite{traceanomaly}.  
In earlier work we had presented values of $\gamma_{\bar\psi\psi,IR}$
\cite{bvh} (see also \cite{ps}) and $\beta'_{IR}$ \cite{bc} calculated as
conventional $n$-loop series expansions in powers of the $n$-loop ($n\ell$) IR
coupling $\alpha_{IR,n\ell}$. In \cite{gtr}-\cite{dexo} we calculated values of
$\gamma_{\bar\psi\psi,IR}$ and $\beta'_{IR}$ via our scheme-independent
expansions. 

In this paper we use our series expansions of $\gamma_{\bar\psi\psi,IR}$ to
$O(\Delta_f^4)$ and of $\beta'_{IR}$ to $O(\Delta_f^5)$ at an IR fixed point in
vectorial, asymptotically free gauge theories with gauge group SU($N_c$) and
$N_f$ Dirac fermions transforming according to various representations $R$, to
calculate Pad\'e approximants to these quantities.  We consider $R$ equal to
the fundamental ($F$), adjoint ($A$) and rank-2 symmetric ($S_2$) tensor
representations. For technical convenience, as before, we restrict to
mass-independent schemes \cite{otherschemes} and zero fermion mass \cite{fm}.
We use these Pad\'e approximants for several purposes, which also constitute
motivations for this work.  First, as closed-form rational functions of
$\Delta_f$, the Pad\'e approximants yield values of $\gamma_{\bar\psi\psi,IR}$
and $\beta'_{IR}$ that complement the values obtained from the finite series
expansions in powers of $\Delta_f$ and can be compared with them.  These values
are of fundamental importance as properties of conformal field theories and are
of value in the resurgent investigation of these theories in four spacetime
dimensions.  Second, our calculations of these quantities in continuum quantum
field theory are complementary to the program of lattice simulations to measure
them, and we compare our values with lattice measurements. Third, given an
asymptotically free theory with a particular choice of gauge group $G$ and
fermion representation $R$, the question of what the infrared properties are,
as a function of $N_f$, is of fundamental field-theoretic importance.  This
applies, in particular, to the question of the extent of the non-Abelian
Coulomb phase as a function of $N_f$. Since the upper end of the NACP at $N_u$
(see Eq. (\ref{nfb1z}) below) \cite{nfintegral} is known exactly, the
determination of the extent of the NACP as a function of $N_f$ is equivalent to
the determination of the value of of $N_f$ at the lower end of the NACP, which
we denote $N_{f,cr}$.  As in our earlier work, here we use our new calculations
of $\gamma_{\bar\psi\psi,IR}$ using Pad\'e approximants, together with an upper
bound on this anomalous dimension from conformal invariance, to obtain further
estimates of $N_{f,cr}$ and hence of the extent of the NACP for several
theories.  Fourth, we carry out Taylor-series expansions of our Pad\'e
approximants in powers of $\Delta_f$ to determine their predictions for
higher-order coefficients in the scheme-independent expansions of
$\gamma_{\bar\psi\psi,IR}$ and $\beta'_{IR}$ going beyond the respective orders
$O(\Delta_f^4)$ and $O(\Delta_f^5)$ to which we have calculated these.  We use
these to test our conjecture in earlier work that the coefficients in the
series expansion of $\gamma_{\bar\psi\psi,IR}$ are positive. (In contrast, we
have shown from our calculations for general $G$ and $R$ in
\cite{dex,dexs,dexl} that the coefficients in the series expansions of
$\beta'_{IR}$ in powers of $\Delta_f$ have mixed signs.)  A fifth use of our
Pad\'e calculations pertains to phenomenology. In addition to the importance of
$N_{f,cr}$ as a basic property describing the UV to IR evolution and infrared
phase structure of an asymptotically free gauge theory, this is also important
for phenomenological studies, since a knowledge of $N_{f,cr}$ is crucial for
the construction of quasi-conformal gauge theories as possible candidates for
ultraviolet completions of the Standard Model.  This is because, for a given
$G$ and $R$, these constructions of quasi-conformal theories require that one
choose $N_f$ to be slightly less than $N_{f,cr}$ in order to achieve the
quasi-conformal behavior whose spontaneous breaking via chiral symmetry
breaking could have the potential to yield a light, dilatonic Higgs-like
scalar.  As a sixth part of the present work, we present a new analytic result
concerning the accuracy of the finite series expansion of
$\gamma_{\bar\psi\psi,IR}$ in powers of $\Delta_f$ as compared with the exactly
known result in an ${\cal N}=1$ supersymmetric gauge theory. This analytical
result extends our earlier demonstrations \cite{gtr,dexs,dexl,dexss} that the
truncation of the series for $\gamma_{\bar\psi\psi,IR}$ to $O(\Delta_f^4)$ is
quite accurate throughout the entire non-Abelian Coulomb phase of this
supersymmetric theory.

This paper is organized as follows. In Section \ref{methods_section} we briefly
review relevant background and methodology. In Sections \ref{gamma_section} and
\ref{betaprime_section} we present our results for $\gamma_{\bar\psi\psi,IR}$
and $\beta'_{IR}$, respectively.  We give our conclusions in Section
\ref{conclusion_section}. 


\section{Background and Methods} 
\label{methods_section} 


\subsection{General}

Here we briefly review some background and methods relevant for our work.  We
refer the reader to our previous papers \cite{gtr}-\cite{dexo} for details.
The requirement of asymptotic freedom implies that $N_f$ must be less than an
upper ($u$) bound $N_u$, where \cite{nfintegral}
\beq
N_u = \frac{11C_A}{4T_f} \ .
\label{nfb1z}
\eeq
Here, $C_2(R)$ is the quadratic Casimir invariant for the representation $R$,
$C_A=C_2(A)$, where $A$ is the adjoint representation, and 
$T_f \equiv T(R)$ is the trace invariant \cite{casimir}.

At the maximal scheme-independent loop order, namely the two-loop level, the 
beta function has an IR zero if $N_f$ lies in the interval $I$ defined by
\beq
I: \quad N_{\ell} < N_f < N_u \ , 
\label{interval}
\eeq
where $N_u$ was given in Eq. (\ref{nfb1z}) and 
\beq
N_{\ell} = \frac{17C_A^2}{4T_f(5C_A+3C_f)} \ .
\label{nfb2z}
\eeq
The maximum value of $\Delta_f$ for $N_f \in I$ is 
\beq
\Delta_{f,max} \equiv N_u - N_\ell = \frac{3C_A(6C_A+11C_f)}{4T_f(5C_A+3C_f)}
\ . 
\label{delta_max}
\eeq

Formally generalizing $N_f$ from positive integers ${\mathbb N}_+$ to positive
real numbers, ${\mathbb R}_+$, one can let $N_f$ approach $N_u$ from below,
thereby making $\alpha_{IR}$ arbitrarily small. Thus, for the UV to IR
evolution in this regime of $N_f$, one infers that the theory evolves from weak
coupling in the UV to an IRFP in a non-Abelian Coulomb phase.  As stated
above, we denote the lowest value of $N_f$ in this NACP as $N_{f,cr}$, and
correspondingly, we define \cite{other} 
\beq
\Delta_{f,cr} \equiv N_u - N_{f,cr} \ . 
\label{delta_cr}
\eeq
so
\beq
{\rm NACP}: \quad N_{f,cr} < N_f < N_u \ .
\label{nacp}
\eeq
One of the goals of lattice studies of these types of gauge theories has been
to estimate $N_{f,cr}$ for a given $G$ and $R$ \cite{lgt}.  We have also
obtained estimates of $N_{f,cr}$ by combining our $O(\Delta_f^4)$ calculations
of $\gamma_{\bar\psi\psi,IR}$ with our finding of monotonicity and the
conformality upper bound \cite{gammabound} $\gamma_{\bar\psi\psi,IR} \le 2$ in
our earlier work \cite{gsi,dex,dexs,dexl}. We will discuss this further below
in connection with our new Pad\'e calculations. Our computations assume that
the IRFP is exact, as is the case in the non-Abelian Coulomb phase
\cite{schisb}. In the analytic expressions and plots given below, this
restriction, that $N_f$ lies in the NACP, will be understood implicitly. In
Table \ref{intervals} we tabulate some relevant values of $N_\ell$, $N_u$,
$\Delta_{f,max}$, and the intervals with $N_f \in {\mathbb R}_+$ and $N_f \in
{\mathbb N}_+$.

For the gauge group $G={\rm SU}(N_c)$ and the specific fermion 
representations $R$ considered in this paper, the general
formulas above for $N_\ell$, $N_u$, and $\Delta_{f,max}$ read as follows:
\beqs
R=F: \quad && N_\ell=\frac{34N_c^3}{13N_c^2-3}, \quad N_u=\frac{11N_c}{2} \ ,
\cr\cr && \Delta_{f,max} = \frac{3N_c(25N_c^2-11)}{2(13N_c^2-3)} \ ,
\label{nuell_fund}
\eeqs
\beqs
R=A: \quad && N_\ell=\frac{17}{16} = 1.0625, \quad N_u=\frac{11}{4} = 2.75 \ ,
\cr\cr && \Delta_{f,max} = \frac{27}{16} = 1.6875 \ ,
\label{nuell_adj}
\eeqs
\beqs
R=S_2: \quad && N_\ell=\frac{17N_c^3}{(N_c+2)(8N_c^2+3N_c-6)}, \cr\cr
             && N_u = \frac{11N_c}{2(N_c+2)} \ , \cr\cr
             && \Delta_{f,max} = \frac{3N_c(18N_c^2+11N_c-22)}
             {2(N_c+2)(8N_c^2+3N_c-6)} \ . \cr\cr
&& 
\label{nuell_sym}
\eeqs
We list these values for the theories under consideration in Table
\ref{intervals}. 


\subsection{Scheme-Independent Expansion for $\gamma_{\bar\psi\psi,IR}$ } 

Since the global chiral symmetry is realized exactly in the non-Abelian Coulomb
phase, the bilinear fermion operators can be classified according to their
representation properties under this symmetry, including flavor-singlet and
flavor-nonsinglet.  The anomalous dimensions evaluated at the IRFP, are the
same for these flavor-singlet and flavor-nonsinglet fermion bilinears
\cite{gracey_op}. For $R=F$, these are $\bar\psi\psi \equiv \sum_{i=1}^{N_f}
\bar\psi_i \psi_i$ and $\bar\psi {\cal T}_b \psi \equiv
\sum_{i,j=1}^{N_f}\bar\psi_i [{\cal T}_b]_{ij} \psi_j$, where ${\cal T}_b$ is a
generator of SU($N_f$)), and similarly for other representations. Hence, as in
our earlier work, we use the symbol $\gamma_{\bar\psi\psi,IR}$ to refer to
both.  This anomalous dimension at the IRFP has the scheme-independent
expansion
\beq
\gamma_{\bar\psi\psi,IR} = \sum_{j=1}^\infty \kappa_j\Delta_f^{\ j} \ . 
\label{gamma_delta_series}
\eeq
We denote the truncation of the infinite series in
Eq. (\ref{gamma_delta_series}) at $j=s$
as$\gamma_{\bar\psi\psi,IR,\Delta_f^s}$.  When it is necessary for clarity, we
indicate the representation $R$ explicitly in the subscripts as
$\gamma_{\bar\psi\psi,IR,R}$, $\gamma_{\bar\psi\psi,IR,R,\Delta_f^s}$, and
$\kappa_{j,R}$.  In general, the calculation of the
coefficient $\kappa_j$ requires, as inputs, the $\ell$-loop coefficients in the
conventional loop expansion of the beta function in powers of $\alpha$, namely 
$b_\ell$, with $1 \le \ell \le j+1$, and the $\ell$-loop
coefficients $c_\ell$ in the corresponding conventional expansion of
$\gamma_{\bar\psi\psi}$ with $1 \le \ell \le j$.  The coefficients
$\kappa_j$ were calculated for general gauge group $G$ and fermion
representation $R$ for $1 \le j \le 3$ in \cite{gtr} and for $j=4$ in
\cite{dexl}.  The calculation of $\kappa_4$ was given for SU(3) and $R=F$ in
\cite{gsi} and for SU($N_c$) in \cite{dexs}, Our calculation of $\kappa_4$ for
general $G$ and $R$ used the $b_\ell$ coefficients up to $b_5$ from \cite{b5}
and $c_\ell$ up to $c_4$ from \cite{c4}. Specific expressions for $\kappa_j$
and plots of $\gamma_{\bar\psi\psi,IR,\Delta_f^s}$ for
$G={\rm SU}(N_c)$ and the fundamental, adjoint, and rank-2 symmetric
and antisymmetric representations were given in \cite{gsi}-\cite{dexl} and
for $G={\rm SO}(N_c)$ and Sp($N_c$) in \cite{dexo}.  In \cite{dex} we discussed
operational criteria for the accuracy of the $\Delta_f$ expansion, and we
briefly review some points here. 

As with usual perturbative series expansions in powers of interaction couplings
in quantum field theories, the $\Delta_f$ expansion is generically an
asymptotic expansion rather than a Taylor series.  This follows from the fact
that in order for an expansion in a variable $z$ to be a Taylor series with
finite radius of convergence, it is necessary (and sufficient) that the
function for which the series is calculated must be analytic at the origin of
the complex $z$ plane.  In particular, with $z = \Delta_f$, this means that,
the properties of the theory should not change qualitatively as one moves from
real positive $\Delta_f$ through the point $\Delta_f=0$ to negative $\Delta_f$,
i.e., as $N_f$ increases through the value $N_u$.  However, one 
knows that the properties of the theory do change qualitatively as $N_f$
increases beyond $N_u$, namely it ceases to be asymptotically free.
Nevertheless, as with conventional perturbative expansions in powers of the
interaction coupling in quantum field theory, one may get a rough estimate of
the accuracy of a truncated series by performing the ratio test on the series
coefficients that have been calculated.  This type of procedure is used, for
example, in perturbative quantum electrodynamics and quantum chromodynamics
calculations in powers of the respective interaction couplings, and we gave
results on the relevant ratios of terms in our series expansions of
$\gamma_{\bar\psi\psi,IR}$ and $\beta'_{IR}$ in our previous work
\cite{gtr}-\cite{dexl}.  These, in conjunction with plots of curves, gave
quantitative evaluations of the accuracy of the $\Delta_f$ expansions for these
quantities. We will expand upon this earlier work here by comparing
$O(\Delta_f^s)$ expansions for $\gamma_{\bar\psi\psi,IR}$ with the exactly
known expression for this anomalous dimension in an ${\cal N}=1$ supersymmetric
gauge theory below. 

Let us denote the full scaling dimension of an operator ${\cal O}$ as $D_{\cal
  O}$ and its free-field value as $D_{{\cal O},free}$. We define the anomalous
dimension of ${\cal O}$, denoted $\gamma_{\cal O}$, by
\beq
D_{\cal O} = D_{{\cal O},free} - \gamma_{\cal O} \ . 
\label{anom_dim}
\eeq
Given that the theory at an IRFP in the non-Abelian phase is conformally
invariant, there is a conformality lower bound on $D_{\cal O}$ from unitarity,
namely $D_{\bar\psi\psi} \ge 1$ \cite{gammabound}. Since
$D_{\bar\psi\psi,free}=3$, this is equivalent to the upper bound
\beq
\gamma_{\bar\psi\psi,IR} \le 2 \ . 
\label{gamma_upperbound}
\eeq
In \cite{gsi}, for SU(3) and $R=F$, using our calculation of
$\gamma_{\bar\psi\psi,IR}$ to $O(\Delta_f^4)$, i.e.,
$\gamma_{\bar\psi\psi,IR,\Delta_f^4}$, we used polynomial extrapolation to
obtain estimates of the evaluation of the infinite series
(\ref{gamma_delta_series}) yielding the value of $\gamma_{\bar\psi\psi,IR}$ as
a function of $N_f$. We compared our results with lattice measurements for 
$N_f=12$ and $N_f=10$. 

In the series expansion (\ref{gamma_delta_series}) for
$\gamma_{\bar\psi\psi,IR}$, the first two coefficients, $\kappa_1$ and
$\kappa_2$, are manifestly positive for any gauge group $G$ and fermion
representation $R$ \cite{gtr}. Although $\kappa_3$ and $\kappa_4$ contain terms
with negative as well as positive signs, one of the important results of our
explicit calculations of $\kappa_3$ and $\kappa_4$ for SU($N_c$), SO($N_c$),
and Sp($N_c$) gauge groups and a variety of representations, including
fundamental, adjoint, and rank-2 symmetric and antisymmetric tensors, was that
for all of these theories, $\kappa_3$ and $\kappa_4$ are also positive
\cite{dex,dexs,dexl,dexo}.  Moreover, as reviewed below, in a gauge theory with
${\cal N}=1$ supersymmetry, an exact expression is known for the anomalous
dimension of the (gauge-invariant) fermion bilinear operator, and the
Taylor-series expansion of this exact expression in powers of $\Delta_f$ yields
$\kappa_j$ coefficients that are all positive.  These results led to our
conjecture in \cite{gsi}, elaborated upon in our later works, that, in addition
to the manifestly positive $\kappa_1$ and $\kappa_2$, and our findings in
\cite{dex,dexs,dexl,dexo} that $\kappa_3$ and $\kappa_4$ are positive for all
of the groups and representations for which we calculated them, (i) the
higher-order $\kappa_j$ coefficients with $j \ge 5$ are also positive in
(vectorial, asymptotically free) non-supersymmetric gauge theories. In turn,
this conjecture led to several monotonicity conjectures, namely that (ii) for
fixed $s$, $\gamma_{\bar\psi\psi,IR,\Delta_f^s}$ increases monotonically as
$N_f$ decreases in the non-Abelian Coulomb phase, and (iii) for fixed $N_f$ in
the NACP, $\gamma_{\bar\psi\psi,IR,\Delta_f^s}$ is a monotonically increasing
function of $s$, so that (iv) for fixed $N_f$ in the NACP and for finite $s$,
$\gamma_{\bar\psi\psi,IR,\Delta_f^s}$ is a lower bound on the actual anomalous
dimension $\gamma_{\bar\psi\psi,IR}$, as defined by the infinite series
(\ref{gamma_delta_series}). By similar reasoning, the analogous conjectures
apply for the $k$'th derivatives of the anomalous dimension as a function of
$\Delta_f$.  In particular, for the first derivative, one has the analogous
conjectures (ii)$_d$ for fixed $s$,
\beq
 \frac{d[\gamma_{\bar\psi\psi,IR,\Delta_f^s}]}{d\Delta_f} = 
-\frac{d[\gamma_{\bar\psi\psi,IR,\Delta_f^s}]}{dN_f} 
\label{derivative_of_gamma}
\eeq
increases monotonically as $\Delta_f$ increases, i.e., as $N_f$ decreases, 
in the NACP (where the subscript $d$ on (ii)$_d$ connotes ``derivative''); 
(iii)$_d$ for fixed $N_f$ in the NACP,
$\frac{d}{d\Delta_f}[\gamma_{\bar\psi\psi,IR,\Delta_f^s}]$ is a monotonically
increasing function of $s$, so that (iv)$_d$ for fixed $N_f$ in the NACP and
for finite $s$, the derivative
$\frac{d}{d\Delta_f}[\gamma_{\bar\psi\psi,IR,\Delta_f^s}]$ is a lower bound on
the derivative $\frac{d}{d\Delta_f}[\gamma_{\bar\psi\psi,IR}]$ of the actual
anomalous dimension, as defined by the infinite series
(\ref{gamma_delta_series}). 

Combining our calculations of $\gamma_{\bar\psi\psi,IR}$ to $O(\Delta_f^4)$
with these positivity and resultant monotonicity conjectures (used as
assumptions), and with the further assumption that $\gamma_{\bar\psi\psi,IR}$
saturates its conformality upper bound of 2 in (\ref{gamma_upperbound}), as
$N_f$ decreases to $N_{f,cr}$ at the lower end of the non-Abelian Coulomb
phase, we have then derived estimates of $N_{f,cr}$ in various theories
\cite{sd}.  For example, in \cite{gsi} we inferred that $N_{f,cr} = 8 - 9$
\cite{gsi,sd}. In \cite{dexs,dexl} we extended these $O(\Delta_f^4)$
calculations of $\gamma_{\bar\psi\psi,IR}$ from the special case of SU(3) and
$R=F$ to general $G$ and $R$, and, for $G={\rm SU}(N_c)$ with various $N_c$ and
$R$, we again compared various our values of
$\gamma_{\bar\psi\psi,IR,\Delta_f^4}$ with values from lattice
measurements. 

Here, we extend this program further via the calculation and
evaluation of Pad\'e approximants for $G={\rm SU}(N_c)$ with several values of
$N_c$ and several fermion representations $R$.  There are a number of
applications of these calculations: (i) to get further estimates of the value
of the anomalous dimension of the fermion bilinear for various $N_c$ and
fermion representations $R$; (ii) to estimate $N_{f,cr}$, as just described;
and (iii) via Taylor series expansions of the Pad\'e approximants, to determine
their predictions for higher-order coefficients $\kappa_j$ with $j \ge 5$.
Of course, regarding application (iii), since the
Pad\'e approximants that we calculate for $\gamma_{\bar\psi\psi,IR}$ are 
based on the series (\ref{gamma_delta_series}) computed only up to 
$O(\Delta_f^4)$, their predictions for these higher-order $\kappa_j$ with 
$j \ge 5$ only provide a hint as to their actual values.


\subsection{Scheme-Independent Expansion for $\beta'_{IR}$ } 

Given the property of asymptotic freedom, $\beta$ is negative in the
region $0 < \alpha < \alpha_{IR}$, and since $\beta$ is continuous, it follows
that this function passes through zero at $\alpha=\alpha_{IR}$ 
with positive slope, i.e., $\beta'_{IR} > 0$. This derivative $\beta'_{IR}$ 
has the scheme-invariant expansion
\beq
\beta'_{IR} = \sum_{j=2}^\infty d_j\Delta_f^{\ j} \ . 
\label{betaprime_delta_series}
\eeq
As indicated, $\beta'_{IR}$ has no term linear in $\Delta_f$.  In general, the
calculation of the scheme-independent coefficient $d_j$ requires, as inputs,
the $\ell$-loop coefficients in the beta function, $b_\ell$, for $1 \le \ell
\le j$.  We denote the truncation of the infinite series in
Eq. (\ref{betaprime_delta_series}) at $j=s$ as
$\beta'_{IR,\Delta_f^s}$. 

Let the full scaling dimension of ${\rm Tr}(F_{\mu\nu}F^{\mu\nu})$ be denoted
$D_{F_{\mu\nu}^2}$ (with free-field value 4). At an IRFP, $D_{F^2,IR} = 4 +
\beta'_{IR}$ \cite{traceanomaly}, so $\beta'_{IR} = -\gamma_{F^2,IR}$. Given
that the theory at an IRFP in the non-Abelian phase is conformally invariant,
there is a conformality bound from unitarity, namely $D_{F^2} \ge 1$
\cite{gammabound}.  Since $\beta'_{IR} > 0$, this bound is obviously satisfied.

As part of our work, we will calculate Pad\'e approximants to our series
expansions to $O(\Delta_f^5)$ for $\beta'_{IR}$.  We will use these for the
analogues of the applications (i) and (iii) mentioned above for
$\gamma_{\bar\psi\psi,IR}$, namely to obtain additional information about the
value of $\beta'_{IR}$ and to get some hints regarding coefficients $d_j$ going
beyond the order to which we have calculated them, i.e., with $j \ge 6$. 


\subsection{LNN Limit}

For $G={\rm SU}(N_c)$ and $R=F$, it is of interest to consider the limit
\beqs
& & LNN: \quad N_c \to \infty \ , \quad N_f \to \infty \cr\cr
& & {\rm with} \ r \equiv \frac{N_f}{N_c} \ {\rm fixed \ and \ finite}  \cr\cr
& & {\rm and} \ \ \xi(\mu) \equiv \alpha(\mu) N_c \ {\rm is \ a \
finite \ function \ of} \ \mu \ .
\cr\cr
& &
\label{lnn}
\eeqs
We will use the symbol $\lim_{LNN}$ for this limit, where ``LNN'' stands
for ``large $N_c$ and $N_f$'' with the constraints in Eq. (\ref{lnn})
imposed.  This is also called the 't Hooft-Veneziano limit. 

Here we give some background for our calculation of Pad\'e approximants in the
LNN limit.  We define
\beq
r_u = \lim_{LNN} \frac{N_u}{N_c} \ ,
\label{rb1zdef}
\eeq
and
\beq
r_\ell = \lim_{LNN} \frac{N_\ell}{N_c} \ ,
\label{rb2zdef}
\eeq
with values
\beq
r_u = \frac{11}{2} =5.5
\label{rb1z}
\eeq
and
\beq
r_\ell = \frac{34}{13}=2.615  
\label{rb2z}
\eeq
(to the indicated floating-point accuracy). With the interval $I$ defined as in
Eq. (\ref{interval}), it follows that the corresponding
interval in the ratio $r$ is
\beq
I_r: \quad \frac{34}{13} < r < \frac{11}{2}, \ i.e., \ 2.615 < r < 5.5 
\label{intervalr}
\eeq
The critical value of $r$ such that for $r > r_{cr}$, the LNN theory is in the
NACP and is IR-conformal, while for $r< r_{cr}$, it exhibits spontaneous chiral
symmetry breaking, is denoted $r_{cr}$ and is defined as
\beq
r_{cr} = \lim_{LNN} \frac{N_{f,cr}}{N_c} \ .
\label{rcr}
\eeq
We define the rescaled scheme-independent expansion 
parameter for the LNN limit
\beq
\Delta_r \equiv \lim_{LNN} \frac{\Delta_f}{N_c} = r_u-r = \frac{11}{2}-r \ .
\label{deltar}
\eeq
As $r$ decreases from $r_u$ to $r_\ell$ in the interval $I_r$,
$\Delta_r$ increases from 0 to a maximal value
\beqs
& & \Delta_{r,max} = r_u-r_\ell = \frac{75}{26} = 2.8846 \quad
{\rm for} \ r \in I_r \ . \cr\cr
& &
\label{deltar_max}
\eeqs
Further, we define the maximum value of $\Delta_r$ in the NACP as 
\beq
\Delta_{r,cr} = r_u - r_{cr} = \frac{11}{2} - r_{cr} \ . 
\label{deltar_cr}
\eeq

Since $\kappa_j \propto N_c^{-j}$, the rescaled coefficients 
$\hat \kappa_{j,F}$ that are finite in the LNN limit are 
\beq
\hat \kappa_{j,F} \equiv \lim_{N_c \to \infty} N_c^j \, \kappa_{j,F} \ . 
\label{kappahatj}
\eeq
The anomalous dimension
$\gamma_{IR}$ is also finite in this limit and is given by
\beq
R=F: \quad \lim_{LNN} \gamma_{IR} = \sum_{j=1}^\infty \kappa_{j,F} \Delta_f^j
= \sum_{j=1}^\infty \hat \kappa_{j,F} \Delta_r^j \ .
\label{gamma_ir_lnn}
\eeq

The appropriately rescaled beta function that is finite in the LNN limit
is
\beq
\beta_\xi = \frac{d\xi}{dt} = \lim_{LNN} N_c \beta \ ,
\label{betaxi}
\eeq
where $\xi$ was defined in Eq. (\ref{lnn}).
Since the derivative $d\beta_\xi/d\xi$ satisfies the relation
\beq
\frac{d \beta_\xi}{d\xi} = \frac{d\beta}{d\alpha} \equiv \beta' \ ,
\label{dbetarelation}
\eeq
it follows that $\beta'$ is finite in the
LNN limit (\ref{lnn}). We define the rescaled coefficient
\beq
\hat d_{j,F} = \lim_{LNN} N_c^j \, d_{j,F} \ ,
\label{dnhat}
\eeq
which is finite in the LNN limit. Thus, writing $\lim_{LNN} \beta'_{IR}$ as
$\beta'_{IR,LNN}$ for this $R=F$ case, we have
\beqs
& & \beta'_{IR,LNN} = \sum_{j=1}^\infty d_{j,F} \Delta_f^j =
\sum_{j=1}^\infty \hat d_{j,F} \Delta_r^j \ . \cr\cr
& &
\label{betaprime_ir_lnn}
\eeqs
We denote the value of $\beta'_{IR,LNN}$ obtained from
this series calculated to order $O(\Delta_f^p)$ as
$\beta'_{IR,LNN,\Delta_f^p}$.


\subsection{Test of Accuracy of Scheme-Independent Expansion for 
$\gamma_{\bar\psi\psi,IR}$ Using Supersymmetric Gauge Theory}

A basic question for the scheme-independent series expansions of physical
quantities at an IRFP in powers of $\Delta_f$ in the non-Abelian Coulomb phase
is how accurate a finite truncation of this series is.  We have addressed this
question in previous work \cite{gtr,dex,dexs,dexl} by investigating how
accurate the truncated, finite-order expansion is, as a function of $N_f$, in a
theory where an exact expression for the anomalous dimension of the fermion
bilinear is known. 

Here we briefly review this analysis and give some new quantitative measures of
the accuracy.  For this accuracy check, we use a vectorial, asymptotically free
gauge theory with ${\cal N}=1$ supersymmetry ($ss$), gauge group $G$, and $N_f$
pairs of chiral superfields $\Phi_j$ and $\tilde\Phi_j$, $j=1,...,N_f$, that
transform according to the respective representations $R$ and $\bar R$ of $G$.
The requirement of asymptotic freedom in this theory requires that $N_f$ must
be less than an upper limit, which we will again denote $N_u$, namely
\beq
ss: \quad N_u = \frac{3C_A}{2T_f} \ . 
\label{nu_susy}
\eeq
(Throughout this subsection, to avoid cumbersome notation, we will use the same
notation $N_u$, $N_\ell$, $N_{f,cr}$, etc. as in the non-supersymmetric case,
but it will be understood implicitly that these quantities refer to this 
supersymmetric theory.)
In this theory, the lower end of the non-Abelian Coulomb phase occurs at
$N_{f,cr}=N_u/2$ , so the NACP occupies the range 
\beqs
ss: \ NACP: \quad && \frac{N_u}{2} < N_f < N_u \ , i.e., \cr\cr
                 && \frac{3C_A}{4T_f} < N_f < \frac{3C_A}{2T_f} \ . 
\label{nacp_susy}
\eeqs
Thus, in this theory, $\Delta_f$ increases from 0 to a maximum value 
\beq
ss: \quad \Delta_{f,max} = \frac{N_u}{2} =  \frac{3C_A}{4T_f} 
\label{delta_max_susy}
\eeq
as $N_f$ decreases from $N_u$ at the upper end of the NACP to $N_u/2$
at the lower end of the NACP.  

The anomalous dimension of the quadratic chiral superfield operator product
$\tilde \Phi \Phi$, and hence also the fermion bilinear contained in this
product, are exactly known in such theories.  Defining the mesonic operator 
$M \equiv \bar\psi\psi$, with a sum over group indices understood, 
one has \cite{nsvz}-\cite{susyreviews}
\beqs
\gamma_{M,IR} &=& \frac{\Delta_f}{N_u-\Delta_f} = 
          \frac{\frac{\Delta_f}{N_u}}{1-\frac{\Delta_f}{N_u}} \cr\cr
     &=& \sum_{j=1}^\infty \bigg ( \frac{\Delta_f}{N_u} \bigg )^j \ . 
\label{gamma_susy}
\eeqs
This anomalous dimension $\gamma_{M,IR}$ increases monotonically from 0 at
$N_f=N_u$ at the upper end of the NACP to saturate its upper limit of 1 when
$N_f$ reaches $N_{f,cr}=N_u/2$ at the lower end of the NACP. From this exact
expression (\ref{gamma_susy}), it follows that the coefficient $\kappa_j$ in
Eq. (\ref{gamma_delta_series}) is
\beq
ss: \quad \kappa_j = \frac{1}{(N_u)^j} = \Big ( \frac{2T_f}{3C_A} \Big )^j \ .
\label{kappaj_ss}
\eeq
Thus, $\kappa_j$ is positive for all $j$, which provided motivation for our
conjecture in \cite{gsi,dex} that $\kappa_j > 0 \ \forall \ j$ in the
non-supersymmetric theory, in accord with the manifestly positive $\kappa_1$
and $\kappa_2$, and the positivity of the $\kappa_j$ with $j=3, \ 4$ that we
had calculated for each group and representation that we considered
\cite{gtr,gsi,dex,dexs,dexl,dexo}.  From this positivity of the $\kappa_j$
calculated to the highest order, $j=4$, the monotonicity property of
$\gamma_{\bar\psi\psi,IR,\Delta_f^4}$ follows.  That is, our calculation of
$\gamma_{\bar\psi\psi,IR,\Delta_f^4}$ in the non-supersymmetric theory shares
with the exact expression in the ${\cal N}=1$ supersymmetric theory the
property that it increases monotonically with decreasing $N_f$ in the NACP.
Note that this monotonicity does not hold for the scheme-dependent conventional
$n$-loop ($n\ell$) calculation of $\gamma_{IR,n\ell}$; for example, it was
found \cite{bvh,ps} that for various specific theories, such as
(non-supersymmetric) SU($N_c$) with $N_c=2, \ 3, \ 4$ and $R=F$, although
$\kappa_3$ is positive in the NACP, $\kappa_4$ (calculated in the widely used
$\overline{\rm MS}$ scheme) is negative, so that, as $N_f$ decreases in the
NACP, $\gamma_{\bar\psi\psi,IR,4\ell}$ reaches a maximum and then decreases.
In the ${\cal N}=1$ supersymmetric theory, we also showed that $\kappa_3$
(calculated in the $\overline{\rm DR}$ scheme) is negative, and, as a
consequence, the scheme-dependent three-loop calculation of
$\gamma_{M,IR,3\ell}$ fails to exhibit the known monotonicity of the exact
result \cite{bfs}.  This again demonstrates the advantage of the
scheme-independent series expansion at the IRFP, (\ref{gamma_delta_series}), in
powers of $\Delta_f$, as compared with conventional expansions in powers of the
coupling $\alpha_{IR}$.

The series expansion for $\gamma_{M,IR}$ is particularly simple in the ${\cal
  N}=1$ supersymmetric gauge theory, since it is a geometric series.  Because
the fermions appear together with the Grassmann variable $\theta$ in the chiral
superfield $\Phi_j = \phi_j + \sqrt{2} \, \theta \psi_j + \theta\theta F_j$
(where $F_j$ is an auxiliary field), the conformality lower bound $D_{\Phi} \ge
1$ on the full scaling dimension of the chiral superfield is equivalent to the
conformality upper bound
\beq
\gamma_{M,IR} \le 1 
\label{gamma_upperbound_susy}
\eeq
on the (gauge-invariant) fermion bilinear in $\Phi_j \tilde \Phi_j$.  This
upper bound is saturated as $N_f$ decreases to $N_{f,cr}$. Thus, as we have
observed before \cite{gtr,dex,dexl,dexss}, both of the assumptions that we make
for our estimate of $N_{f,cr}$ in non-supersymmetric theories, namely that (i)
$\kappa_j > 0$ for all $j$, and (ii) $\gamma_{\bar\psi\psi,IR}$ saturates its
upper bound from conformal invariance as $N_f$ decreases to the lower end of
the non-Abelian Coulomb phase, are satisfied in a gauge theory with ${\cal
  N}=1$ supersymmetry. (As discussed above, the upper bounds themselves are
different, namely 2 in the non-supersymmetric theory,
Eq. (\ref{gamma_upperbound}) and 1 in the supersymmetric theory,
Eq. (\ref{gamma_upperbound_susy}) \cite{nonsat}.)

We next determine the accuracy of a finite truncation of the series
(\ref{gamma_delta_series}). To do this we calculate the fractional difference 
\beq
\epsilon_{ss} \equiv 
\frac{\gamma_{M,IR}-\gamma_{M,IR,\Delta_f^s}}{\gamma_{M,IR}} \ , 
\label{epsilon_susy}
\eeq
where we denote the finite series (\ref{gamma_delta_series}) for 
$\gamma_{M,IR}$ truncated to maximal power $j=s$ as 
$\gamma_{M,IR,\Delta_f^s}$. Using the elementary identity $\sum_{j=1}^s x^j =
x(x^s-1)/(x-1)$ to sum the finite series $\sum_{j=1}^s \kappa_j \Delta_f^j$, 
we obtain
\beq
\gamma_{M,IR,\Delta_f^s} = \frac{\Big ( \frac{\Delta_f}{N_u} \Big ) 
\Big [ \Big ( \frac{\Delta_f}{N_u}\Big )^s - 1 \Big ]}
{\frac{\Delta_f}{N_u} - 1} \ . 
\label{gammas_susy}
\eeq
Substituting this into Eq. (\ref{epsilon_susy}), we find
\beq
\epsilon_{ss} = \Big ( \frac{\Delta_f}{N_u} \Big )^s \ . 
\label{epsilon_susy_value}
\eeq
Since the maximum value of the ratio $\Delta_f/N_u$ is 1/2, this fractional
difference decreases toward zero exponentially rapidly with $s$.
Quantitatively, if one sets $N_f=N_u/2$, the value at the bottom of the NACP,
then $\epsilon_{ss}=(1/2)^s = e^{-(\ln 2)s}$, so the fractional difference
between the $O(\Delta_f^4)$ result, $\gamma_{M,IR,\Delta_f^4}$, and the exact
result, $\gamma_{M,IR}$, is 6.25 \%. If one formally sets $N_f=(3/4)N_u$,
further up in the NACP, then $\Delta_f = N_u/4$ and the fractional difference
between $\gamma_{M,IR,\Delta_f^s}$ and the exact result is
$\epsilon_{ss}=(1/4)^s$. This takes on the value 0.391 \% for $s=4$.  In
general, if we require that $\epsilon_{ss} < \epsilon_0$ for some $\epsilon_0 >
0$, then this implies that it is necessary to calculate the finite, truncated
series in powers of $\Delta_f$ up to and including the power
\beq
s = \frac{\ln \Big ( \frac{1}{\epsilon_0} \Big )}
{\ln \Big ( \frac{N_u}{\Delta_f} \Big ) } \ , 
\label{smin_for_epsilon}
\eeq
to achieve this fractional accuracy, where it is understood that if $s$ is a
non-integral real number, then one sets $s$ equal to the closest integer
greater than the value in Eq. (\ref{smin_for_epsilon}).  At the upper end of
the non-Abelian Coulomb phase, since $\Delta_f/N_u$ is small,
$\ln(N_u/\Delta_f)$ is large and one can achieve a small fractional difference
$\epsilon_0$ with a modest value of $s$.  The most stringent requirement on $s$
to achieve a given fractional accuracy $\epsilon_0$ occurs as $N_f$ 
approaches the lower end of the NACP at $N_f=N_u/2$ and is 
\beq
s = \frac{\ln \Big ( \frac{1}{\epsilon_0} \Big )}{\ln 2} \ . 
\label{s_at_lower_end_nacp}
\eeq
As we have calculated, if one wants to achieve a fractional difference that is
less than or equal to 6.25 \% for all $N_f$ in the NACP, then
Eq. (\ref{smin_for_epsilon}) shows that the expansion to $O(\Delta_f^4)$ is
sufficient for this accuracy.

These results show quantitatively that finite truncations of the
infinite series (\ref{gamma_delta_series}), even up to only modest maximal
powers such as $s=4$, yield very accurate approximations to the exactly known
anomalous dimension $\gamma_{M,IR}$ in this ${\cal N}=1$ supersymmetric gauge
theory.  In passing, we remark that if an exact expression for $\beta'_{IR}$
were available in this theory, then we could also use it to obtain an
additional measure of how accurate a finite-order truncation of the infinite
series (\ref{betaprime_delta_series}) is to the exact function.  However, to
our knowledge, an exact expression for $\beta'_{IR}$ is not known for this
theory.

This ${\cal N}=1$ supersymmetric gauge theory also provides a framework in
which to investigate how accurate a $[p,q]$ Pad\'e approximant to a
finite-order truncation of the infinite series in Eq. (\ref{gamma_susy}) would
be to the exact result in (\ref{gamma_susy}) for $\gamma_{M,IR}$.  In general,
if one calculates a $[p,q]$ Pad\'e approximant for a finite truncation of such
a simple series as the geometric series in (\ref{gamma_susy}), then not only
does the [0,1] approximant reproduce the exact function (\ref{gamma_susy}), but
so do all of the $[p,q]$ approximants with $q \ne 0$.  The way that they do
this is by inserting factors in the numerator and denominator to yield
polynomials of degree $p$ and $q$, but which cancel precisely, yielding the
exact function (\ref{gamma_susy}) itself.

From the exact expression (\ref{gamma_susy}), one can also calculate the
value of the derivative $d\gamma_{M,IR}/dN_f$, which is 
\beq
\frac{d\gamma_{M,IR}}{dN_f} = - \frac{d\gamma_{M,IR}}{d\Delta_f} = 
-\frac{N_u}{N_f^2} \ . 
\label{dgamma_susy_dNf}
\eeq
This derivative is always negative in the NACP and increases monotonically in
magnitude with decreasing $N_f$. It has the value $-1/N_u$ for
$N_f=N_u$ at the upper end of the NACP, and $-4/N_u$ for $N_f=N_u/2$ at the 
lower end of the NACP. The curvature is
\beq
\frac{d^2\gamma_{M,IR}}{dN_f^2} = \frac{d^2\gamma_{M,IR}}{d\Delta_f^2} = 
\frac{2N_u}{N_f^3} \ . 
\label{gamma_susy_curvature}
\eeq
This curvature is positive in the NACP and increases from $2/N_u^2$ at the
upper end, to $16/N_u^2$, at the lower end, of the NACP.  Given that we have
shown that $\kappa_1$ and $\kappa_2$ are manifestly positive and $\kappa_j$ are
positive for all the $G$ and $R$ for which we have evaluated them, it follows
that our $\gamma_{\bar\psi\psi,IR,\Delta_f^4}$ also has positive curvature for
$N_f$ in the NACP, a property that it shares with the exactly known
$\gamma_{M,IR}$ in the ${\cal N}=1$ supersymmetric theory. We showed in
\cite{bfs} that the three-loop calculation of
$\gamma_{M,IR,3\ell}$ in the supersymmetric gauge theory, carried out as a 
conventional scheme-dependent series expansion in powers of $\alpha$, 
fails to exhibit the known positive curvature of the exact result, just as it
fails to exhibit the known monotonicity of the exact result.  This is another
advantage of the scheme-independent expansion in powers of $\Delta_f$. 

We have noted above that the value of $N_f$ at the lower end of the
NACP does not, in general, coincide with the value $N_\ell$ at the lower end of
the interval $I$ where the two-loop beta function has an IR zero. For this
${\cal N}=1$ supersymmetric theory, one can calculate this difference exactly
\cite{cfwindow,bfs}.  One has 
\beq
N_{\ell} = \frac{3C_A^2}{2T_f(C_A+2C_f)} \ . 
\label{nell_susy}
\eeq
Hence, this difference for this theory is 
\beq
N_{\ell} - N_{f,cr} = \frac{3C_A(C_A-2C_f)}{4T_f(C_A+2C_f)} 
\label{nfdiff_susy}
\eeq
(where here and in the rest of this subsection, $N_\ell$ is given by
(\ref{nell_susy}) and should not be confused with Eq. (\ref{nfb2z})).  
This difference can be positive or negative. 
For example, for $G={\rm SU}(N_c)$ and $R=F$, this difference is the positive
quantity 
\beq
R=F: \quad N_{\ell} - N_{f,cr} = \frac{3N_c}{2(2N_c^2-1)} \ . 
\label{nfdiff_fund_susy}
\eeq
This decreases to zero as $N_c \to \infty$. The resultant fractional 
difference between these values is
\beq
R=F: \quad \frac{N_{\ell} - N_{f,cr}}{N_{\ell}} = \frac{1}{2N_c^2-1} \ . 
\label{ndiff_fund_su_fractional}
\eeq
This fractional difference (\ref{ndiff_fund_su_fractional}) has the values
0.143 and 0.0588 for $N_c=2$ and $N_c=3$, respectively, and also decreases
toward zero with increasing $N_c$.  In contrast, for both the adjoint and
symmetric rank-2 tensor representations, the difference
$N_\ell-N_{f,cr}$ is negative and does not vanish as $N_c \to \infty$ (with 
$N_f$ fixed). 

In the LNN limit of the ${\cal N}=1$ supersymmetric gauge theory with 
$G={\rm SU}(N_c)$ and $R=F$,
\beq
LNN: \quad r_{cr} = r_\ell = \frac{3}{2} \ , 
\label{rcr_lnn_susy}
\eeq
so that 
\beq
LNN: \quad \Delta_{r,cr}=\Delta_{r,max}= \frac{3}{2} \ .
\label{deltamax_lnn_susy} 
\eeq
%


\section{Pad\'e Approximants for $\gamma_{\bar\psi\psi,IR}$ using the 
$O(\Delta_f^4)$ Series} 
\label{gamma_section}

\subsection{General} 

In this section we report our calculation of Pad\'e approximants to our
scheme-independent $O(\Delta_f^4)$ series for $\gamma_{\bar\psi\psi,IR}$, in
SU($N_c$) theories with various fermion representations $R$. (For a review of
Pad\'e approximants, see, e.g., \cite{padereview}.) It may be recalled that
resummation methods such as Pad\'e approximants have been useful in 
in the analysis of series expansions in both quantum field theories 
and in statistical mechanics \cite{genpade} (e.g. \cite{bl}-\cite{bgn}), 
and our current work extends to higher order our earlier calculations of Pad\'e
approximants for these types of gauge theories \cite{bvh,flir,gsi,dex}. 

In general, given the series calculated to maximal order $s$, denoted
$\gamma_{\bar\psi\psi,IR,\Delta_f^s}$ as above (with $N_c$ and $R$ implicit in
the notation), we write this as
\beqs
\gamma_{\bar\psi\psi,IR,\Delta_f^s} &=&
\sum_{j=1}^s \kappa_j \Delta_f^j \cr\cr
&=&  \kappa_1 \Delta_f \Big [ 
1 + \frac{1}{\kappa_1} \sum_{j=2}^s \kappa_j \Delta_f^{j-1} \Big ] 
\label{gamma_reduced}
\eeqs
and calculate the $[p,q]$ Pad\'e approximant to the expression in square
brackets, with $p+q=s-1$. This takes the form 
\beq
\gamma_{\bar\psi\psi,IR,[p,q]} = 
\kappa_1 \Delta_f \bigg [ \frac{1 + \sum_{i=1}^p {\cal N}_i \Delta_f^i}
{1 + \sum_{j=1}^q {\cal D}_j \Delta_f^j} \bigg ] \ , 
\label{gamma_pqpade}
\eeq
where 
\beq
\kappa_1 = \frac{8C_fT_f}{C_A(7C_A+11C_f)} \ .
\label{kappa1}
\eeq
The $[p,q]$ Pad\'e approximant is thus a rational function whose $p+q$
coefficients are determined uniquely by the condition that the Taylor series
expansion of this approximant must match the $s-1$ coefficients in the series
in square brackets.  With the prefactor $\kappa_1 \Delta_f$ thus extracted, the
Pad\'e approximant in the square brackets is normalized to be equal to 1 at
$\Delta_f=0$. By construction, the series expansion of each closed-form
approximant $\gamma_{\bar\psi\psi,IR,[p,q]}$ exactly reproduces the series
expansion of $\gamma_{\bar\psi\psi,IR}$ up to the maximal order to which we
have calculated it, $s=4$.  In addition to providing a closed-form
rational-function approximation to the finite series, a Pad\'e approximant also
can be used in another way, namely to yield a hint of higher-order terms.  This
information is obtained by carrying this Taylor series expansion of $[p,q]$
Pad\'e approximants with $q \ne 0$ to higher order.  
We will use the Pad\'e approximants for both of these applications. 

Note that the $[s-1,0]$ Pad\'e
approximant is just the series itself, i.e.,
\beq
\gamma_{\bar\psi\psi,IR,[s-1,0]} = \gamma_{\bar\psi\psi,IR, \Delta_f^s} \ . 
\label{gammajm1pade_relation}
\eeq
As a special case of Eq. (\ref{gammajm1pade_relation}) for
$j_{max}=4$, $\gamma_{\bar\psi\psi,IR,[3,0]}$ is just the original polynomial 
$\gamma_{\bar\psi\psi,IR,\Delta_f^4}$ itself, and hence we do not
consider it, since we have already obtained evaluations of this truncated
series in previous work. 

By construction, the $[p,q]$ Pad\'e approximant in (\ref{gamma_pqpade}) is
analytic at $\Delta_f=0$, and if it has $q \ne 0$, then it is a meromorphic
function with $q$ poles.  A necessary condition that must be satisfied for a
Pad\'e approximant to be useful for our analysis here is that it must not have
a pole for any $\Delta_f$ in the interval $0 < \Delta_f < \Delta_{cr}$, or
equivalently, for any $N_f$ in the non-Abelian Coulomb phase. Since $N_{f,cr}$
is not precisely known for all values of $N_c$ and all fermion representations
$R$ under consideration here, we will also use another condition, namely that
the Pad\'e approximant should not have any poles for $N_f$ in the interval $I$
where the two-loop beta function has an IR zero, or equivalently, for
$\Delta_f$ in the interval $0 < \Delta_f < \Delta_{f,max}$.  This second
condition can be applied in a straightforward manner for each $N_c$ and $R$,
since the upper and lower ends of this interval $I$, namely $N_u$ and $N_\ell$,
and thus $\Delta_{f,max}$, are known (listed above in Eqs. (\ref{nfb1z}),
(\ref{nfb2z}), and (\ref{delta_max})). Since the $[p,q]$ Pad\'e
approximant in Eq. (\ref{gamma_pqpade}) is an analytic function at
$\Delta_f=0$, and, for $q \ne 0$, is meromorphic, the radius of convergence of
its Taylor series expansion is set by the magnitude of the pole closest to the
origin in the complex $\Delta_f$ plane. Let us denote this radius of
convergence as $\Delta_{f,conv.}$.  We shall also require that
$\Delta_{f,conv.}$ be greater than $\Delta_{f,max}$ and $\Delta_{f,cr}$, since
we would like the Taylor series expansion of (\ref{gamma_pqpade}) to accurately
reproduce the series (\ref{gamma_delta_series}) in this disk.  We will do this
to be as careful as possible, even though the actual expansion
(\ref{gamma_delta_series}) is, in general, only expected to be an asymptotic
expansion. 


\subsection{$G={\rm SU}(N_c)$, $R=F$}

With $G={\rm SU}(N_c)$ and $R=F$, the explicit expression for $\kappa_{1,F}$ is
\beq
\kappa_{1,F} = \frac{4(N_c^2-1)}{N_c(25N_c^2-1)} \ .
\label{kappa1_fund}
\eeq
The explicit numerical expressions for the scheme-independent series expansions
of $\gamma_{\bar\psi\psi,IR}$ to order $\Delta_f^4$ for $R=F$ and $N_c=2, \ 3,
\ 4$ are as follows:
\begin{widetext}
\beqs
{\rm SU}(2): \ \gamma_{\bar\psi\psi,IR,F,\Delta_f^4} & = &
   0.0674157 \Delta_f 
+ (0.733082 \times 10^{-2}) \Delta_f^2
+ (0.605308 \times 10^{-3}) \Delta_f^3
+ (1.626624 \times 10^{-4}) \Delta_f^4 \ , 
\cr\cr
& &
\label{gamma_Delta_p4_su2}
\eeqs
\beqs
{\rm SU}(3): \ \gamma_{\bar\psi\psi,IR,F,\Delta_f^4} & = &
   0.0498442 \Delta_f
+ (3.79282 \times 10^{-3}) \Delta_f^2
+ (2.37475 \times 10^{-4}) \Delta_f^3
+ (3.67893 \times 10^{-5}) \Delta_f^4 \ , 
\cr\cr
& &
\label{gamma_Delta_p4_su3}
\eeqs
\beqs
{\rm SU}(4): \ \gamma_{\bar\psi\psi,IR,F,\Delta_f^4} & = &
   0.0385604 \Delta_f
+ (2.231420 \times 10^{-3}) \Delta_f^2
+ (1.122984 \times 10^{-4}) \Delta_f^3
+ (1.265054 \times 10^{-5}) \Delta_f^4 \ . 
\cr\cr
& &
\label{gamma_Delta_p4_su4}
\eeqs
\end{widetext}
In these equations,
\beq
\Delta_f = \frac{11N_c}{2} - N_f \quad {\rm for} \ R=F \ .
\label{Deltaf_fund}
\eeq
For reasons of space, in 
Eqs. (\ref{gamma_Delta_p4_su2})-(\ref{gamma_Delta_p4_su4}) we list the
expansions to the given floating-point accuracy; our actual algebraic computer 
calculations of Pad\'e approximants use the coefficients in these expansions to
considerably higher numerical accuracy.) 


\subsection{SU(2)}

\begin{figure}
  \begin{center}
    \includegraphics[height=6cm]{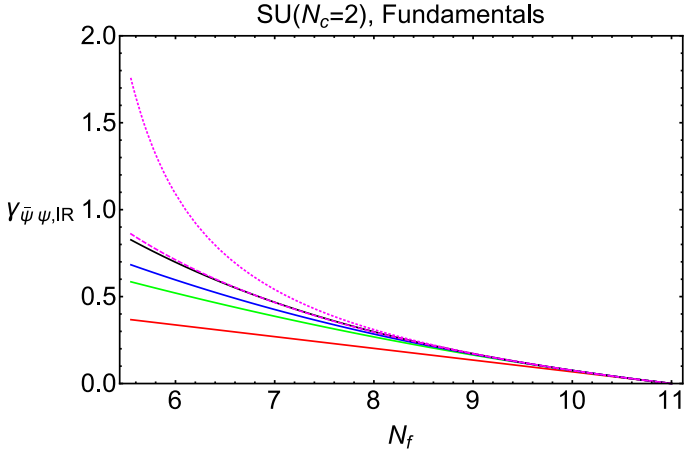}
  \end{center}
\caption{Plot of $\gamma_{\bar\psi\psi,IR,F,[1,2]}$ and
                 $\gamma_{\bar\psi\psi,IR,F,[0,3]}$ for $N_c=2$, i.e., 
SU(2) and $R=F$, together with 
$\gamma_{\bar\psi\psi,IR,F,\Delta_f^s}$ with $1 \le s \le 4$ and 
the extrapolated value 
$\gamma_{\bar\psi\psi,IR,F} = \lim_{s \to \infty}
\gamma_{\bar\psi\psi,IR,F,\Delta_f^s}$, as a function 
of $N_f$. The vertical axis is labelled generically as
$\gamma_{\bar\psi\psi,IR}$. 
From bottom to top, the curves (with colors online) refer to 
$\gamma_{\bar\psi\psi,IR,F,\Delta_f}$ (red),
$\gamma_{\bar\psi\psi,IR,F,\Delta_f^2}$ (green),
$\gamma_{\bar\psi\psi,IR,F,\Delta_f^3}$ (blue), and
$\gamma_{\bar\psi\psi,IR,F,\Delta_f^4}$ (black). The curves for 
$\gamma_{\bar\psi\psi,IR,F,[1,2]}$ and
$\gamma_{\bar\psi\psi,IR,F,[0,3]}$ are dashed magenta and dotted magenta. 
Over most of the range of $N_f$, the dashed curve for the extrapolated value 
$\gamma_{\bar\psi\psi,IR,F}$ lies highest with cyan color online.}
\label{gammaNc2fund_plot}
\end{figure}

For $G={\rm SU}(2)$ with $R=F$, the general formulas (\ref{nfb1z}) and
(\ref{nfb2z}) give $N_u=11$ and $N_\ell=5.551$, so the interval $I$ with $N_f
\in {\rm \mathbb R}_+$ is $5.551 < N_f < 11$, with $\Delta_{f,max} = N_u-N_\ell
= 5.449$, and the physical interval with $N_f \in {\mathbb N}_+$ is $6 \le N_f
\le 10$. This information is summarized in Table \ref{intervals}. For this
SU(2) theory with $R=F$, we calculate the following $[p,q]$ Pad\'e approximants
with $q \ne 0$:
\begin{widetext}
\beq
\gamma_{\bar\psi\psi,F,[2,1]} = 0.0674157\Delta_f \Bigg [
\frac{ 1 -0.159986\Delta_f - 0.0202427 \Delta_f^2}{1-0.268727 \Delta_f} 
\Bigg ] \ , 
\label{gamma_fund_su2_pade21}
\eeq
\beq
\gamma_{\bar\psi\psi,F,[1,2]} = 0.0674157\Delta_f \Bigg [ 
\frac{1 + 0.613518\Delta_f }{1+0.504778\Delta_f-0.0638685\Delta_f^2} \Bigg ] \
, 
\label{gamma_fund_su2_pade12}
\eeq
\beq
\gamma_{\bar\psi\psi,F,[0,3]} = 0.0674157\Delta_f \Bigg [ 
\frac{1}{1-0.1087405\Delta_f+(2.845756 \times 10^{-3})\Delta_f
-(1.745923 \times 10^{-3})\Delta_f^3} \Bigg ] \ .
\label{gamma_fund_su2_pade03}
\eeq
\end{widetext}
The [2,1] Pad\'e approximant in $\gamma_{\bar\psi\psi,F,[2,1]}$ has a pole at
$\Delta_f = 3.721$ i.e., at $N_f=7.279$.  This is in the interval $I$ and in
the estimated NACP, so we cannot use this [2,1] Pad\'e approximant for our
analysis.  The [1,2] Pad\'e approximant in $\gamma_{\bar\psi\psi,F,[1,2]}$ has
poles at $\Delta_f = -1.6405$, i.e., $N_f=12.6405$, and at $\Delta_f = 9.544$,
i.e., $N_f=1.456$. The first of these occurs at a value of $N_f$ greater than
$N_u$, while the second occurs at a value of $N_f$ well below $N_\ell$ and
$N_{f,cr}$, so neither is in the interval $I$ or in the NACP.  Finally, the
[0,3] Pad\'e approximant has a pole at $\Delta_f=6.289$, i.e., at $N_f=4.731$,
which is below $N_\ell$ and slightly below the estimated $N_{f,cr}$.  In
addition, this [0,3] approximant has a complex-conjugate pair of poles at
$\Delta_f = -2.3195 \pm 9.273i$, whose magnitude is 9.558, considerably greater
than $\Delta_{f,max}=5.449$ and $\Delta_{cr} \simeq 5.5$. Hence, we can use the
[1,2] and [0,3] Pad\'e approximants for our analysis. In Table
\ref{gamma_fund_values} we list values of $\gamma_{\bar\psi\psi,IR,F,[1,2]}$
and $\gamma_{\bar\psi\psi,IR,F,[0,3]}$ for this SU(2) theory with $R=F$ as a
function of $N_f$ and, for comparison, values of
$\gamma_{\bar\psi\psi,IR,F,\Delta_f^s}$ for $1 \le s \le 4$ from
\cite{dex,dexl}. In Fig. \ref{gammaNc2fund_plot} we plot these values.

We next make some general comments about these SU(2), $R=F$ calculations, which
also will apply to our calculations for SU(3) and SU(4) with $R=F$.  In earlier
work \cite{gtr}-\cite{dexo}, we have noted that, in addition to the manifestly
positive $\kappa_1$ and $\kappa_2$, the higher-order coefficients $\kappa_j$
with $j=3$ and $j=4$ are positive for all of the groups, SU($N_c$), SO($N_c$),
and Sp($N_c$) and for all of the representations, namely $F$, $A$, $S_2$, and
$A_2$, for which we have performed these calculations.  Here (at an IR fixed
point in the non-Abelian Coulomb phase), this means that,
at least with $s$ in the range $1 \le s \le 4$, 
(i) for fixed $s$, $\gamma_{\bar\psi\psi,IR,F,\Delta_f^s}$ monotonically
increases with decreasing $N_f$; and (ii) for fixed $N_f$ and hence $\Delta_f$,
$\gamma_{\bar\psi\psi,IR,F,\Delta_f^s}$ is a monotonically increasing function
of $s$. As is evident from Table \ref{gamma_fund_values} and
Fig. \ref{gammaNc2fund_plot}, the analogue of the monotonicity property (i) is
also true for the Pad\'e approximants, namely that both the [1,2] and [0,3]
Pad\'e approximants increase monotonically with decreasing $N_f$ values listed
in the table and shown in the figure.  The value of the anomalous dimension
obtained via the [1,2] Pad\'e approximant, $\gamma_{\bar\psi\psi,IR,F,[1,2]}$,
is quite close to $\gamma_{\bar\psi\psi,IR,F,\Delta_f^4}$, increasing slightly
above it as $N_f$ decreases toward the lower part of the non-Abelian Coulomb
phase. The curve for $\gamma_{\bar\psi\psi,IR,F,[0,3]}$ lies above that for
$\gamma_{\bar\psi\psi,IR,F,[1,2]}$ and increases more rapidly with decreasing
$N_f$.  

In \cite{dexs,dexl} we compared our results for
$\gamma_{\bar\psi\psi,IR,F,\Delta_f^s}$ with $s$ up to 4 in this SU(2) theory
with our earlier conventional $n$-loop calculations in \cite{bvh} and with
lattice measurements for $N_f=8$ \cite{su2nf8,ckm}. These lattice measurements
are consistent with the SU(2), $R=F$, $N_f=8$ theory being IR-conformal, so our
calculations in the non-Abelian Coulomb phase are applicable. As listed in
Table \ref{gamma_fund_values}, rounding off to two significant figures, we have
$\gamma_{\bar\psi\psi,IR,F,\Delta_f^3} = 0.29$ and
$\gamma_{\bar\psi\psi,IR,F,\Delta_f^4} = 0.30$, in very good agreement with our
two Pad\'e values, $\gamma_{\bar\psi\psi,IR,F,[1,2]} = 0.30$ and
$\gamma_{\bar\psi\psi,IR,F,[0,3]} = 0.31$ and slightly higher than our earlier
$n$-loop calculations $\gamma_{\bar\psi\psi,IR,F,3\ell}=0.27$ and
$\gamma_{\bar\psi\psi,IR,F,4\ell}=0.20$.

We now go further to combine our scheme-independent series calculation of
$\gamma_{\bar\psi\psi,IR,\Delta_f^4}$ with our Pad\'e approximant computation
to estimate $N_{f,cr}$ for this theory.  We require that both of
the two Pad\'e-based values, namely $\gamma_{\bar\psi\psi,IR,F,[1,2]}$ and
$\gamma_{\bar\psi\psi,IR,F,[0,3]}$, should obey the conformality upper bound
(\ref{gamma_upperbound}). We also assume that the larger of these values
saturates this upper bound at the lower end of the NACP, as the exactly known
$\gamma_{M,IR}$ saturates its upper bound in the supersymmetric gauge theory
discussed above.  Then the $N_f$ value at which the larger of these two values
exceeds the conformality upper bound yields the new Pad\'e-based estimate of
$N_{f,cr}$.  For this SU(2) theory (and for the SU(3) and SU(4) theories to be
discussed below) the larger Pad\'e-based value is
$\gamma_{\bar\psi\psi,IR,F,[0,3]}$.  From Fig.  \ref{gammaNc2fund_plot}, we
therefore infer that
\beq
{\rm SU}(2), \ R=F: \quad N_{f,cr} \simeq 5-6 \ . 
\label{nfcr_fund_su2}
\eeq
This result is consistent with a recent lattice study \cite{su2nf6} of the
SU(2) theory with $R=F$, $N_f=6$, which finds this theory is
IR-conformal. (Earlier lattice studies of this theory include
\cite{su2nf6_earlier}.)  With this estimate that $N_{f,cr} \lsim 6$, it follows
that the non-Abelian Coulomb phase occupies the physical interval $6 \le N_f
\le 10$, the same as the interval $I$ with $N_f \in {\mathbb N}_+$.

As another application, we can calculate Taylor series expansions of these
Pad\'e approximants to see what they predict for higher-order coefficients,
namely the $\kappa_{j,F}$ with $j \ge 5$. (Recall that the $\kappa_{j,R}$ were
given for general $G$ and $R$ in \cite{dex,dexs,dexl} and were listed
numerically for $G={\rm SU}(2)$, $R=F$ in Eq.  (\ref{gamma_Delta_p4_su2})
above.)  We find that the Pad\'e approximant $\gamma_{\bar\psi\psi,IR,F,[0,3]}$
that we used to estimate $N_{f,cr}$ yields coefficients $\kappa_{j,F}$ with $j
\ge 5$ that are all positive to the highest order to which we have calculated
them, namely $j=200$. This is an important result, because it shows that our
use of this approximant, $\gamma_{\bar\psi\psi,IR,F,[0,3]}$, to estimate
$N_{f,cr}$ is self-consistent.  That is, our use assumed that, in addition to
the known positive $\kappa_{j,F}$ with $1 \le j \le 4$, the higher-order
$\kappa_j$ with $j \ge 5$ are positive, and our Taylor series expansion of
$\gamma_{\bar\psi\psi,IR,F,[0,3]}$ is consistent with this.  In Table
\ref{kappaj_pade_fund} we list the higher-order coefficients
$\kappa_{j,F,[0,3]}$ with $5 \le j \le 10$ for this theory obtained from the
Taylor series expansions of the Pad\'e approximant
$\gamma_{\bar\psi\psi,IR,F,[0,3]}$.  Since we did not use the
$\gamma_{\bar\psi\psi,IR,F,[2,1]}$ for our estimate of $N_{f,cr}$, it is not of
direct relevance what the signs of the $\kappa_{j,F}$ with $j \ge 5$ from the
Taylor series expansion of this [2,1] approximant.  However, for completeness,
we mention that they include both positive and negative ones in an alternating
manner.  The first few are $\kappa_{5,F,[1,2]}=-(0.4344482 \times 10^{-4})$,
$\kappa_{6,F,[1,2]}= 0.323207 \times 10^{-4}$, $\kappa_{7,F,[1,2]}=-(1.90897
\times 10^{-5})$, etc.  This difference in the signs of the
$\kappa_{j,F,[0,3]}$ and $\kappa_{j,F,[1,2]}$ for $j \ge 5$ accounts for the
fact that $\gamma_{\bar\psi\psi,IR,F,[0,3]} >
\gamma_{\bar\psi\psi,IR,F,[1,2]}$, as observed in Table \ref{gamma_fund_values}
and Fig. \ref{gammaNc2fund_plot}.  Similar comments apply for the SU(3) and
SU(4) theories with $R=F$ to be discussed next.


\subsection{SU(3)}

For $G={\rm SU}(3)$ with $R=F$, the general formulas (\ref{nfb1z}) and
(\ref{nfb2z}) yield the values $N_u=16.5$, $N_\ell=8.053$.  Thus, for this
theory, the interval $I$ for $N_f \in {\mathbb R}_+$ is $8.053 < N_f < 16.5$
with $\Delta_{f,max}=8.447$, and the physical interval $I$ with $N_f \in
{\mathbb N}_+$ is $9 \le N_f \le 16$ (see Table \ref{intervals}).  
From our calculation of $\gamma_{\bar\psi\psi,IR,F,\Delta_f^4}$ in \cite{gsi},
we presented polynomial extrapolations to infinite order to obtain estimates of
$\lim_{s \to \infty} \gamma_{\bar\psi\psi,IR,\Delta_f^s}$. 
Combining these with the conformality upper
bound (\ref{gamma_upperbound}) and the assumption that, as in the
supersymmetric case, $\gamma_{\bar\psi\psi,IR}$ saturates this upper bound at
the lower end of the NACP, we estimated that \cite{gsi}
\beq
{\rm SU}(3), \ R=F: \quad N_{f,cr} \simeq 8-9 \ , 
\label{nfcr_fund_su3}
\eeq
in agreement with several lattice estimates \cite{lgt},
\cite{lsd}-\cite{lmnp}. As we will discuss, our new calculations presented here
are consistent, to within the intrinsic theoretical uncertainties involved,
with our estimate of $N_{f,cr}$ given in \cite{gsi}.

For this SU(3) theory, we calculate the following $[p,q]$ Pad\'e approximants
with $q \ne 0$:
\begin{widetext}
\beq
\gamma_{\bar\psi\psi,F,[2,1]} = 0.0498442\Delta_f \Bigg [
\frac{ 1 -0.0788254\Delta_f - 0.00702398 \Delta_f^2}{1-0.154919\Delta_f} 
\Bigg ] \ , 
\label{gamma_fund_su3_pade21}
\eeq
\beq
\gamma_{\bar\psi\psi,F,[1,2]} = 0.0498442\Delta_f \Bigg [ 
\frac{1 + 0.442170\Delta_f }{1+0.366077\Delta_f-0.0326204\Delta_f^2} \Bigg ] 
\ , 
\label{gamma_fund_su3_pade12}
\eeq
\beq
\gamma_{\bar\psi\psi,F,[0,3]} = 0.0498442\Delta_f \Bigg [ 
\frac{1}{1-0.0760935\Delta_f+(1.02588 \times 10^{-3})\Delta_f
-(0.4536614 \times 10^{-3})\Delta_f^3} \Bigg ] \ .
\label{gamma_fund_su3_pade03}
\eeq
\end{widetext}
The [2,1] Pad\'e approximant in $\gamma_{\bar\psi\psi,F,[2,1]}$ has a pole at
$\Delta_f = 6.455$, i.e., at $N_f=10.045$. Hence, this pole lies in the
interval $I$ and also in the NACP, so we cannot use this [2,1] Pad\'e
approximant for our analysis.  The [1,2] Pad\'e approximant in
$\gamma_{\bar\psi\psi,F,[1,2]}$ has poles at $\Delta_f = -2.272$, i.e.,
$N_f=18.772$, and at $\Delta_f = 13.494$, i.e., $N_f=3.006$. The first of these
occurs at a value of $N_f$ greater than $N_u$, while the second occurs at a
value of $N_f$ well below $N_\ell$ and $N_{f,cr}$, so neither is in the
interval $I$ or in the NACP.  Finally, the [0,3] Pad\'e approximant has a pole
at $\Delta_f=9.392$, i.e., $N_f=7.108$, and a complex-conjugate pair of poles
at $\Delta_f = -3.565 \pm 14.900i$.  The real pole lies below $N_\ell$ and the
above-mentioned estimates of the lower end of the NACP at $N_f =8-9$, while the
complex poles have magnitude 15.321, which is considerably larger than
$\Delta_{f,max}=8.447$ and $\Delta_{cr} \simeq 8$. Hence, we can make use of
both the [1,2] and [0,3] Pad\'e approximants for our analysis.

\begin{figure}
  \begin{center}
    \includegraphics[height=6cm]{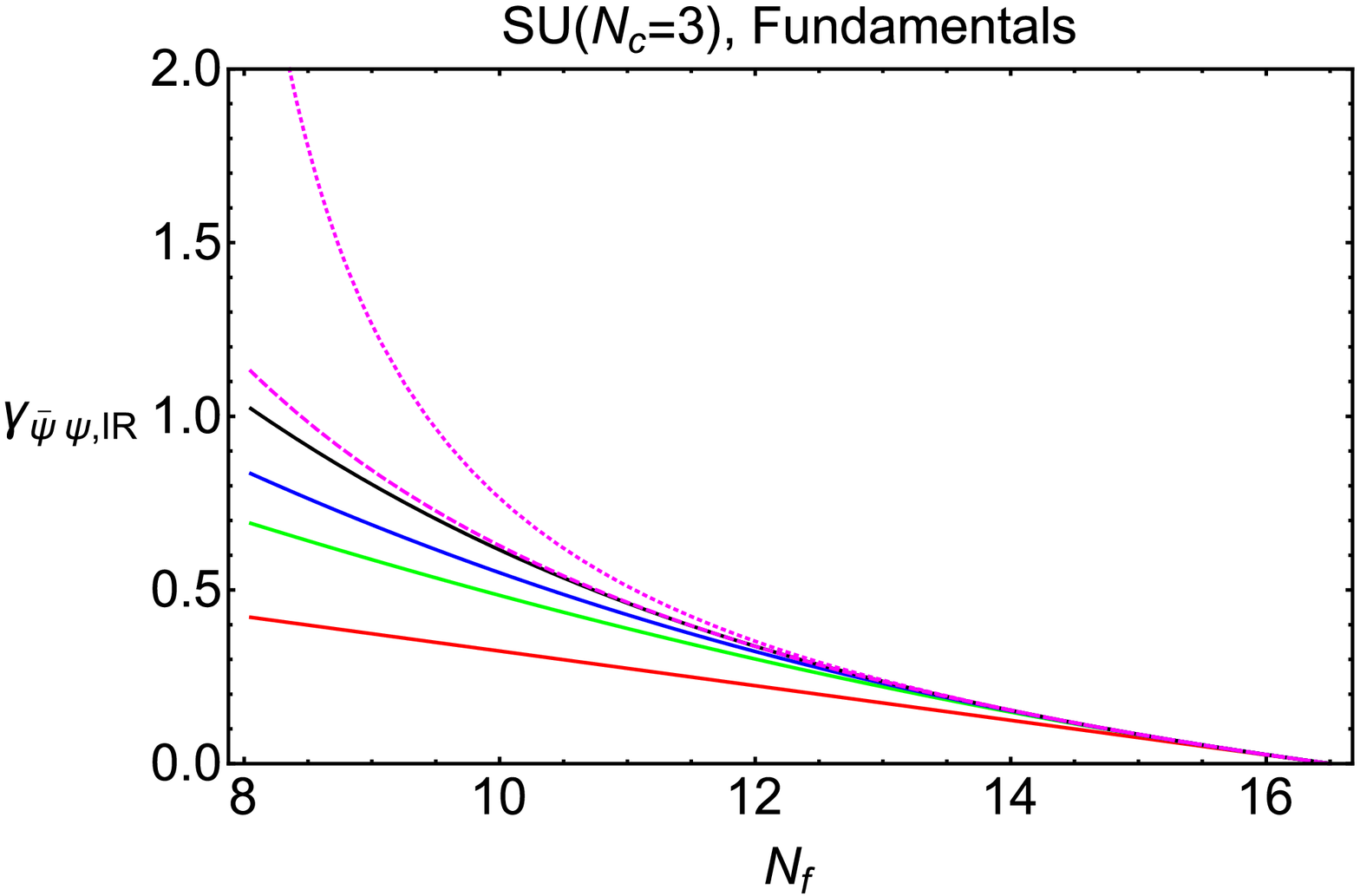}
  \end{center}
\caption{Plot of $\gamma_{\bar\psi\psi,IR,F,[1,2]}$ and
                 $\gamma_{\bar\psi\psi,IR,F,[0,3]}$ for $N_c=3$, i.e., 
SU(3) and $R=F$, together with 
$\gamma_{\bar\psi\psi,IR,F,\Delta_f^s}$ with $1 \le s \le 4$ and 
the extrapolated value 
$\gamma_{\bar\psi\psi,IR,F} = \lim_{s \to \infty}
\gamma_{\bar\psi\psi,IR,F,\Delta_f^s}$, as a function 
of $N_f$. The vertical axis is labelled generically as
$\gamma_{\bar\psi\psi,IR}$. 
From bottom to top, the curves (with colors online) refer to 
$\gamma_{\bar\psi\psi,IR,F,\Delta_f}$ (red),
$\gamma_{\bar\psi\psi,IR,F,\Delta_f^2}$ (green),
$\gamma_{\bar\psi\psi,IR,F,\Delta_f^3}$ (blue), and
$\gamma_{\bar\psi\psi,IR,F,\Delta_f^4}$ (black). The curves for 
$\gamma_{\bar\psi\psi,IR,F,[1,2]}$ and
$\gamma_{\bar\psi\psi,IR,F,[0,3]}$ are dashed magenta and dotted magenta. 
Over most of the range of $N_f$, the dashed curve for the extrapolated value 
$\gamma_{\bar\psi\psi,IR,F}$ lies highest with cyan color online.}
\label{gammaNc3fund_plot}
\end{figure}

In Table \ref{gamma_fund_values} we list values of
$\gamma_{\bar\psi\psi,IR,F,[1,2]}$ and $\gamma_{\bar\psi\psi,IR,F,[0,3]}$ for
this SU(3) theory as a function of $N_f$ and, for comparison, values
of $\gamma_{\bar\psi\psi,IR,F,\Delta_f^s}$ for $1 \le s \le 4$ from
\cite{dex,dexl}. In Fig. \ref{gammaNc3fund_plot} we plot
these two Pad\'e approximants to our $O(\Delta_f^4)$ series,
$\gamma_{\bar\psi\psi,IR,F,[1,2]}$ and $\gamma_{\bar\psi\psi,IR,F,[0,3]}$,
as a function of $N_f$ for $N_f \in I$. For
comparison, we also show our previously calculated
$\gamma_{\bar\psi\psi,IR,F,\Delta_f^s}$ with $1 \le s \le 4$. 

The general features that we remarked on for the SU(2) theory with $R=F$ are
also evident here.  For $N_f$ in the upper part of the non-Abelian Coulomb
phase, the values of $\gamma_{\bar\psi\psi,IR,F,[1,2]}$ and
$\gamma_{\bar\psi\psi,IR,F,[0,3]}$ are quite close to
$\gamma_{\bar\psi\psi,IR,R,\Delta_f^4}$.  As $N_f$ decreases,
$\gamma_{\bar\psi\psi,IR,F,[1,2]}$ continues to be close to
$\gamma_{\bar\psi\psi,IR,F,\Delta_f^4}$, while
$\gamma_{\bar\psi\psi,IR,R,[0,3]}$ becomes progressively larger than
$\gamma_{\bar\psi\psi,IR,F,\Delta_f^4}$ and $\gamma_{\bar\psi\psi,IR,F,[1,2]}$.
For small $N_f$ near to the lower end of the non-Abelian Coulomb phase,
$\gamma_{\bar\psi\psi,IR,F,[0,3]}$ rises up and eventually exceeds the
conformality upper bound (\ref{gamma_upperbound}) for $N_f$ between 8 and 9.
Using the value of $N_f$ where $\gamma_{\bar\psi\psi,IR,F,[0,3]}$ exceeds this
upper bound as an estimate of $N_{f,cr}$, we derive the result $N_{f,cr} \sim
8-9$, in agreement with Eq. (\ref{nfcr_fund_su3}) from \cite{gsi} and with most
lattice estimates.

Our new results from the Pad\'e approximants extend our previous comparison
with lattice measurements in \cite{gsi,dex,dexs,dexl}. There have been a number
of studies of the SU(3) theory with $R=F$ and $N_f=12$, including
\cite{lsd}-\cite{kuti}. Although many lattice groups have concluded that
this theory is IR-conformal, there is not yet a consensus on this point (for 
recent reviews, see \cite{lgt}).  As listed in Table \ref{gamma_fund_values},
rounding off to two significant figures, for this SU(3) theory with $R=F$ and 
$N_f=12$, our new values from the Pad\'e approximants, namely
$\gamma_{\bar\psi\psi,IR,F,[1,2]} = 0.34$ and $\gamma_{\bar\psi\psi,IR,F,[0,3]}
= 0.35$, are in good agreement with our scheme-independent series calculations
in \cite{gsi}, namely, $\gamma_{\bar\psi\psi,IR,F,\Delta_f^3} = 0.32$ and
$\gamma_{\bar\psi\psi,IR,F,\Delta_f^4} = 0.34$, and are somewhat higher than
our conventional $n$-loop calculations,
$\gamma_{\bar\psi\psi,IR,F,3\ell}=0.31$,
$\gamma_{\bar\psi\psi,IR,F,4\ell}=0.25$ \cite{bvh}, and
$\gamma_{\bar\psi\psi,IR,F,5\ell}=0.26$ \cite{gsi}.  The lattice simulations
have obtained a range of values for $\gamma_{\bar\psi\psi,IR,F}$,
including the following: $\gamma_{\bar\psi\psi,IR,F} \sim 0.414(16)$
\cite{lsd}, $\gamma_{\bar\psi\psi,IR,F} \simeq 0.35$ \cite{degrand},
$\gamma_{\bar\psi\psi,IR,F} \simeq 0.4$ \cite{latkmi},
$\gamma_{\bar\psi\psi,IR,F} = 0.27(3)$ \cite{ah1}, $\gamma_{\bar\psi\psi,IR,F}
\simeq 0.25$ \cite{ah2} (see also \cite{ah3}), $\gamma_{\bar\psi\psi,IR,F} =
0.235(46)$ \cite{lmnp}, and $0.2 \lsim \gamma_{\bar\psi\psi,IR,F} \lsim 0.4$
\cite{kuti}.  Here, the quoted uncertainties in the last digits
are indicated in parentheses; we refer the reader to these papers for detailed
discussions of overall uncertaintites in these measurements. Recent critical
discussions of these lattice measurements include \cite{lgt,kuti,arw}.

It is also worthwhile to compare our results with other lattice studies,
bearing in mind that (i) our calculations assume an exact IR fixed point, as is
true in the non-Abelian Coulomb phase, and (ii) as $N_f$ decreases toward the
lower end of the NACP and $\Delta_f$ increases, one generally needs more terms
in a series expansion in powers of $\Delta_f$ to achieve a given accuracy.  For
SU(3), $R=F$, and $N_f=10$ (see Table \ref{gamma_fund_values}), again rounding
off to two significant figures, we obtain the Pad\'e approximant values
$\gamma_{\bar\psi\psi,IR,F,[1,2]} = 0.63$ and $\gamma_{\bar\psi\psi,IR,F,[0,3]}
= 0.76$.  The first of these Pad\'e-based values is close to our highest-order
scheme-independent series calculation \cite{gsi},
$\gamma_{\bar\psi\psi,IR,F,\Delta_f^4} = 0.62$, while the second is slightly
higher.  A study of the SU(3) theory with $R=F$ and $N_f=10$ was reported in
\cite{su3nf10}, with the result $\gamma_{\bar\psi\psi,IR,F} \sim O(1)$.  To
within the estimated uncertainties, our values for this theory from \cite{gsi},
as augmented by our new results from Pad\'e approximants, are in reasonable
agreement with this estimate of $\gamma_{\bar\psi\psi,IR,F}$ from
\cite{su3nf10}.  For SU(3) with $R=F$ and $N_f=8$, as is evident in Table
\ref{gamma_fund_values}, there is a significant difference between the values
of our two Pad\'e approximants, indicating that a calculation of the series to
higher order in $\Delta_f$ than $O(\Delta_f^4)$ would be desirable.  This
theory with $N_f=8$ been the subject of a number of lattice studies, including
\cite{latkminf8,lsdnf8}, which have observed quasi-conformal behavior. However,
there has not yet been a decisive conclusion on whether or
not this SU(3) theory with $R=F$ and $N_f=8$ is in the (chirally symmetric) 
non-Abelian Coulomb phase, where the IRFP is exact and our calculations apply,
or in the chirally broken phase \cite{schisb}. 

The Taylor series expansion of the Pad\'e approximant
$\gamma_{\bar\psi\psi,IR,F,[0,3]}$ to calculate higher-order coefficients
$\kappa_{j,F}$ with $j \ge 5$ is again of interest for this SU(3) theory. We
list the higher-order coefficients $\kappa_{j,F}$ with $5 \le j \le 10$ from
the Taylor series expansion of $\gamma_{\bar\psi\psi,IR,F,[0,3]}$ in Table
\ref{kappaj_pade_fund}.  As was the case with the SU(2) theory, these
coefficients are all positive.  We find the same positivity for the highest
order, $j=200$, to which we have calculated the series expansion of this Pad\'e
approximant in powers of $\Delta_f$.  As with SU(2), this shows the
self-consistency of our use of $\gamma_{\bar\psi\psi,IR,F,[0,3]}$ here to
estimate $N_{f,cr}$, which assumed this positivity of higher-order
coefficients.  Similarly to the SU(2) theory, the higher-order coefficients
$\kappa_{j,F}$ from the Taylor series expansion of the other approximant,
$\gamma_{\bar\psi\psi,IR,F,[1,2]}$, which we did not use to estimate $N_{f,cr}$
(since it is smaller than use of $\gamma_{\bar\psi\psi,IR,F,[0,3]}$) have
alternating signs, starting with a negative $\kappa_{5,F}$.


\subsection{SU(4)}

For SU(4) with $R=F$, the general formulas (\ref{nfb1z}) and (\ref{nfb2z})
yield the values $N_u=22$, $N_\ell=10.615$.  Thus, for this theory, the
interval $I$ with $N_f \in {\mathbb R}_+$ is $10.615 < N_f < 22$, with 
$\Delta_{f,max}=11.385$, and the physical interval $I$ with $N_f \in {\mathbb
  N}_+$ is $11 \le N_f \le 21$. 

For this SU(4) theory, we calculate the following $[p,q]$ Pad\'e approximants
with $q \ne 0$:
\begin{widetext}
\beq
\gamma_{\bar\psi\psi,F,[2,1]} = 0.0385604\Delta_f \Bigg [
\frac{ 1 -0.0547830\Delta_f - (0.360664 \times 10^{-2})\Delta_f^2}
{1-0.112651\Delta_f} 
\Bigg ] \ , 
\label{gamma_fund_su4_pade21}
\eeq
\beq
\gamma_{\bar\psi\psi,F,[1,2]} = 0.0385604\Delta_f \Bigg [ 
\frac{1 + 0.423414\Delta_f }{1+0.365546\Delta_f-0.0240657\Delta_f^2} \Bigg ] 
\ , 
\label{gamma_fund_su4_pade12}
\eeq
\beq
\gamma_{\bar\psi\psi,F,[0,3]} = 0.0385604\Delta_f \Bigg [ 
\frac{1}{1-0.05786815\Delta_f+(0.436451 \times 10^{-3})\Delta_f
-(1.847995 \times 10^{-4})\Delta_f^3} \Bigg ] \ .
\label{gamma_fund_su4_pade03}
\eeq
\end{widetext}
The [2,1] Pad\'e approximant in $\gamma_{\bar\psi\psi,F,[2,1]}$ has a pole at
$\Delta_f = 8.877$, i.e., at $N_f=13.123$.  Since this pole lies in the
interval $I$, we do not use the [2,1] Pad\'e approximant.  The [1,2] Pad\'e
approximant in $\gamma_{\bar\psi\psi,F,[1,2]}$ has poles at $\Delta_f =
-2.367$, i.e., $N_f=24.367$, and at $\Delta_f = 17.556$, i.e., $N_f=4.444$. The
first of these occurs at a value of $N_f$ greater than $N_u$, while the second
occurs at a value of $N_f$ well below $N_\ell$. Finally, the [0,3] Pad\'e has a
pole at $\Delta_f=12.379$, i.e., $N_f=9.621$, which is below $N_\ell$. This
[0,3] approximant also has complex poles at $\Delta_f = -5.0085 \pm 20.299i$,
of magnitude 20.908, which is considerably greater than $\Delta_{f,max}=11.385$
and $\Delta_{f,cr} \simeq 11$.  We can thus use both the [1,2] and [0,3] Pad\'e
approximants for our analysis.  This is the same set that we could use in the
case of the SU(2) and SU(3) gauge theories with $R=F$.

\begin{figure}
  \begin{center}
    \includegraphics[height=6cm]{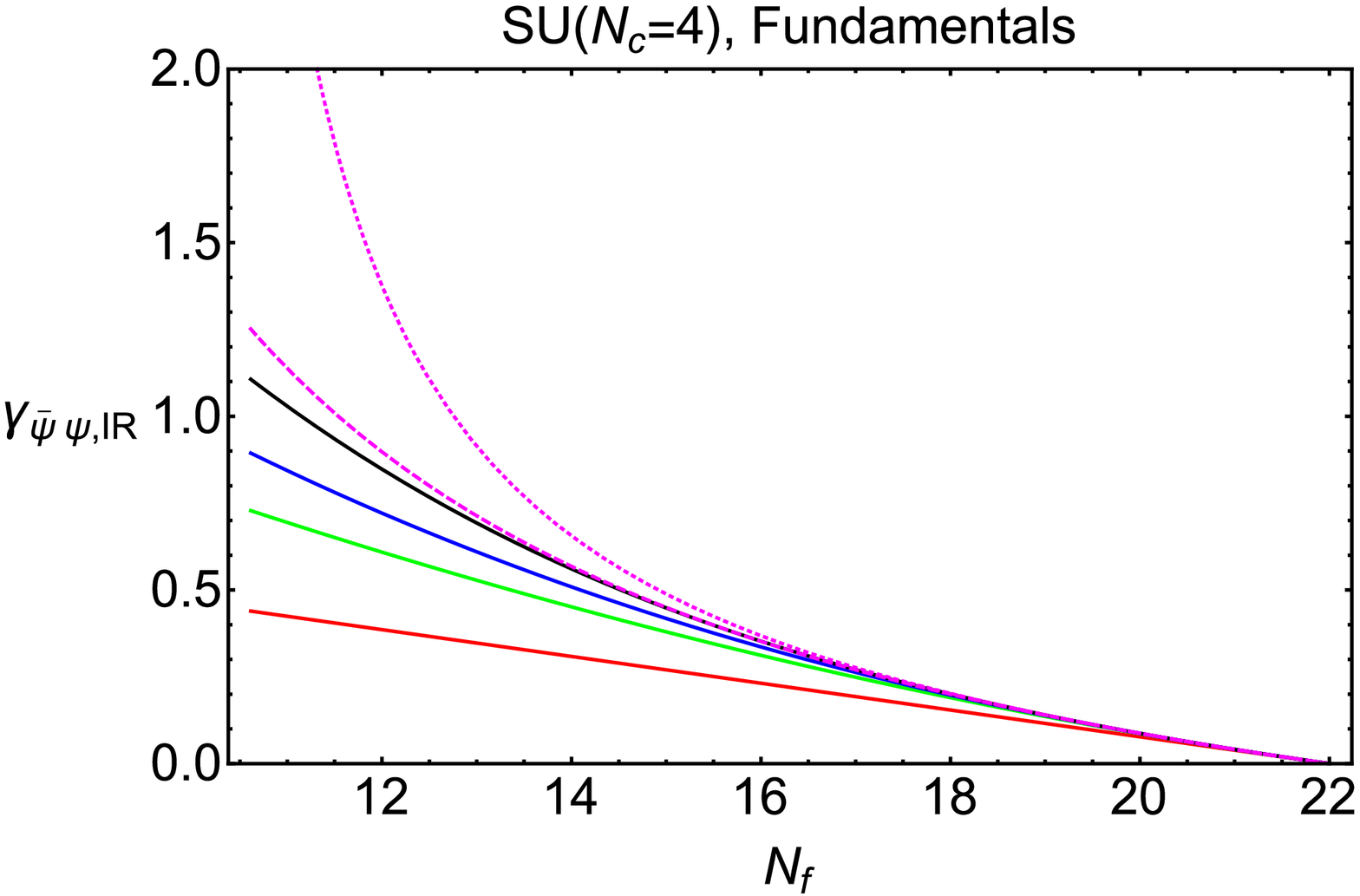}
  \end{center}
\caption{Plot of $\gamma_{\bar\psi\psi,IR,F,[1,2]}$ and
                 $\gamma_{\bar\psi\psi,IR,F,[0,3]}$ for $N_c=4$, i.e., 
SU(4) and $R=F$, together with 
$\gamma_{\bar\psi\psi,IR,F,\Delta_f^s}$ with $1 \le s \le 4$ and 
the extrapolated value 
$\gamma_{\bar\psi\psi,IR,F} = \lim_{s \to \infty}
\gamma_{\bar\psi\psi,IR,F,\Delta_f^s}$, as a function 
of $N_f$. The vertical axis is labelled generically as
$\gamma_{\bar\psi\psi,IR}$. 
From bottom to top, the curves (with colors online) refer to 
$\gamma_{\bar\psi\psi,IR,F,\Delta_f}$ (red),
$\gamma_{\bar\psi\psi,IR,F,\Delta_f^2}$ (green),
$\gamma_{\bar\psi\psi,IR,F,\Delta_f^3}$ (blue), and
$\gamma_{\bar\psi\psi,IR,F,\Delta_f^4}$ (black). The curves for 
$\gamma_{\bar\psi\psi,IR,F,[1,2]}$ and
$\gamma_{\bar\psi\psi,IR,F,[0,3]}$ are dashed magenta and dotted magenta. 
Over most of the range of $N_f$, the dashed curve for the extrapolated value 
$\gamma_{\bar\psi\psi,IR,F}$ lies highest with cyan color online.}
\label{gammaNc4fund_plot}
\end{figure}

In Table \ref{gamma_fund_values} we list the values of the Pad\'e
approximants $\gamma_{\bar\psi\psi,IR,F,[1,2]}$ and $\gamma_{\bar\psi\psi,IR,F,
  [0,3]}$ for this SU(4) theory, as a function of $N_f$.  For
comparison, we also include the values of
$\gamma_{\bar\psi\psi,IR,F,\Delta_f^s}$ with $1 \le s \le 4$ from
\cite{dex,dexl}. In Fig. \ref{gammaNc4fund_plot} we plot all of these values. 
The general features of these results for SU(4) are similar to the features
that we have already discussed for SU(2) and SU(3).  Again using 
the larger of the two Pad\'e-based values, $\gamma_{\bar\psi\psi,IR,F,[0,3]}$,
and inferring $N_{f,cr}$ as the value of $N_f$ where this exceeds the
conformality upper bound, we derive the estimate 
\beq
{\rm SU}(4): \quad N_{f,cr} \sim 11 \ .
\label{nfcr_fund_su4}
\eeq
The self-consistency of our procedure is again shown by the fact that
the higher-order coefficients $\kappa_{j,F}$ with $j \ge 5$ from the Taylor
series expansion of $\gamma_{\bar\psi\psi,IR,F,[0,3]}$ are positive. We list
these $\kappa_{j,F}$ in Table \ref{kappaj_pade_fund}.


%


\subsection{ LNN Limit} 

Here, in the LNN limit, we calculate Pad\'e approximants for our series
$\gamma_{\bar\psi\psi,IR,\Delta_r^4}$ from $s=4$ in \cite{dex,dexl}. 
The values of $\hat\kappa_{j,F}$ were given (analytically) 
in \cite{dex} for $1 \le j \le 3$ and in \cite{dexl} for $j=4$, and are 
\beq
\hat\kappa_{1,F} = \frac{2^2}{5^2} = 0.1600 \ ,
\label{kappahat1}
\eeq
\beq
\hat\kappa_{2,F} = \frac{588}{5^6} = 0.037632 \ ,
\label{kappahat2}
\eeq
\beq
\hat\kappa_{3,F} =
\frac{2193944}{3^3 \cdot 5^{10}} = 0.83207 \times 10^{-2} \ ,
\label{kappahat3}
\eeq
\beqs
\hat\kappa_{4,F} & = & \frac{210676352}{3^4 \cdot 5^{13}}
+ \frac{90112}{3^3 \cdot 5^{10}} \zeta_3
+ \frac{11264}{3^3 \cdot 5^8} \zeta_5 \cr\cr
& = & 0.36489 \times 10^{-2} \ .
\label{kappahat4}
\eeqs
The resultant $\gamma_{\bar\psi\psi,IR,F,LNN,\Delta_r^4}$ is
\beqs
&& {\rm LNN}, \ R=F: \cr\cr
&& \gamma_{\bar\psi\psi,IR,F,LNN,\Delta_r^4} =
0.160000\Delta_f + 0.0376320 \Delta_r^2 \cr\cr
&&+(0.832074 \times 10^{-2})\Delta_r^3 
  +(0.364894 \times 10^{-2})\Delta_r^4 \ . \cr\cr
&&
\label{gamma_ir_lnn_Delta4}
\eeqs
In Eq. (\ref{gamma_ir_lnn_Delta4}) we have listed
$\gamma_{\bar\psi\psi,IR,F,LNN,\Delta_r^4}$ in numerical form, to the indicated
floating-point accuracy, but in our actual computer calculations, the
coefficients are used to considerably higher accuracy.)

Analogously to Eq. (\ref{gamma_reduced}), we write 
\beqs
\gamma_{\bar\psi\psi,IR,F,LNN,\Delta_r^s} &=&
\sum_{j=1}^s \hat\kappa_{j,F} \Delta_r^j \cr\cr
&=&  \hat \kappa_{1,F} \Delta_r \Big [ 
1 + \frac{1}{\hat\kappa_{1,F}} \sum_{j=2}^s \hat\kappa_{j,F} 
\Delta_r^{j-1} \Big ] \cr\cr
&&
\label{gamma_reduced_lnn}
\eeqs
and calculate the $[p,q]$ Pad\'e approximant to the expression in square
brackets, with $p+q=s-1$.  We have calculated analytic results for Pad\'e
approximants to $\gamma_{\bar\psi\psi,IR,F,\Delta_r^4}$. It is again simplest
to present these in numerical form.

We find 
\begin{widetext} 
\beq
\gamma_{\bar\psi\psi,IR,F,LNN,[2,1]} = \frac{4}{25}\Delta_r \Bigg [ 
\frac{1 -0.203335\Delta_r -0.0511389\Delta_r^2}{1 - 0.438535\Delta_r} \Bigg ] 
\ , 
\label{gamma_lnn_pade21}
\eeq
\beq
\gamma_{\bar\psi\psi,IR,F,LNN,[1,2]} = \frac{4}{25}\Delta_r \Bigg [ 
\frac{1 + 3.425595\Delta_r}{1 + 3.190395\Delta_r -0.80238556\Delta_r^2} \Bigg ]
\ , 
\label{gamma_lnn_pade12}
\eeq
\beq
\gamma_{\bar\psi\psi,IR,F,LNN,[0,3]} = \frac{4}{25}\Delta_r \Bigg [ 
\frac{1}{1 - 0.235200\Delta_r + 0.00331444\Delta_r^2 -0.0113539\Delta_r^3} 
\Bigg ] \ . 
\label{gamma_lnn_pade03}
\eeq
\end{widetext}
The [2,1] Pad\'e approximant has a pole at $\Delta_r = 2.28032$, or
equivalently, $r=3.21968$, which lies in the interval $I_r$ and also in the
inferred NACP (see Eq. (\ref{rcr} below)), and hence 
we cannot use this approximant for our analysis. The [1,2] Pad\'e approximant
has poles at $\Delta_r= -0.291997$ and $\Delta_r=4.26813$, i.e., at $r=5.79200$
and $r=1.23187$.  The first pole lies above $r_u$, where the theory is not
asymptotically free, and the second pole lies below $r_\ell$.  
Considering $\gamma_{\bar\psi\psi,IR,F,LNN,[1,2]}$ as an analytic function, the
pole at $\Delta_r=-0.291997$ lies much closer to the origin $\Delta_r=0$ than
the radii of both of the disks $|\Delta_r | < \Delta_{r,max}=2.8846$ and 
$|\Delta_r | < \Delta_{r,cr}=2.6$ (where $\Delta_{r,cr}$ is given below in
Eq. (\ref{deltar_cr})). Consequently, we cannot use this [1,2] Pad\'e
fully reliably for our analysis.  However, it turns out that because the pole
lies on the opposite side of the origin in the $\Delta_r$ plane, at negative
$\Delta_f$, relative to the interval $I: \ 0 < \Delta_r < 2.8846$ and the
NACP, where $0 < \Delta_r \lsim 2.9$, the approximant
$\gamma_{\bar\psi\psi,IR,F,LNN,[1,2]}$ is actually rather close to 
$\gamma_{\bar\psi\psi,IR,F,\Delta_r^4}$. The [0,3] Pad\'e approximant 
has a zero at $\Delta_r=3.03365$, which is larger than $\Delta_{r,max}$ and
$\Delta_{r,cr}$.  In terms of $r$, this pole is at $r=2.46635$, which lies 
below $r_\ell$.  This [0,3] approximant also has 
complex poles at $\Delta_r = -1.370865 \pm 5.210899i$ with magnitude 
$|\Delta_r|=5.38820$, which is larger than the values 
$\Delta_{r,max}=2.8846$ and $\Delta_{r,cr}=2.6$. 
Therefore, we can use the [0,3] Pad\'e approximant reliably. 

In Table \ref{gamma_lnn_values} we list values of
$\gamma_{\bar\psi\psi,IR,F,[1,2]}$ and 
$\gamma_{\bar\psi\psi,IR,F,[0,3]}$ for this
LNN limit and, for comparison, values of 
$\gamma_{\bar\psi\psi,IR,F,\Delta_r^s}$
for $1 \le s \le 4$ from \cite{dex,dexl}.  We have remarked above that
although one of the poles in $\gamma_{\bar\psi\psi,IR,F,[1,2]}$ lies within the
disk $|\Delta_r| < \Delta_{r,max}$ in the complex $\Delta_r$ plane, this pole
does not occur in the region of positive $\Delta_r$ of interest here and hence
does not strongly affect the values of 
$\gamma_{\bar\psi\psi,IR,F,[1,2]}$ in this region. 
As was the case with the specific
SU($N_c$) theories with $R=F$ discussed above, for each of the given $r$ values
in Table \ref{gamma_lnn_values}, $\gamma_{\bar\psi\psi,IR,F,[0,3]}$ is larger
than $\gamma_{\bar\psi\psi,IR,F,[1,2]}$. At the upper end of the NACP, these
are very close to each other, as they are to 
$\gamma_{\bar\psi\psi,F,\Delta_r^4}$.  As $r$ decreases sufficiently, 
$\gamma_{\bar\psi\psi,IR,F,[0,3]}$ first exceeds the conformality upper bound
at $r \simeq 2.9$.  Thus, if we use this value as the estimate of the lower end
of the NACP, we infer that
\beq
LNN: \quad r_{cr} \simeq 2.9
\label{rcr_lnn}
\eeq
and hence 
\beq
LNN: \quad \Delta_{r,cr} \simeq 2.6 \ . 
\label{delta_cr_lnn}
\eeq
Our inferred value of $r_{cr}$ defining the lower end of the NACP is slightly
larger than the value of $r$ defining the lower end of the interval $I$,
namely, $r_\ell=2.615$, and correspondingly, the maximal value of $\Delta_r$ in
the NACP, $\Delta_{r,cr} \simeq 2.6$, is slightly smaller than the maximal
value of $\Delta_r$ in the interval $I$, namely $\Delta_{r,max}=2.8846$, as
given in (\ref{deltar_max}).

It is of interest to investigate how close the $N_{f,cr}$ values inferred for
SU($N_c$) theories with $R=F$ and finite $N_c$ and $N_f$ are to our result
(\ref{rcr_lnn}) in the LNN limit.  To make this comparison, we compute
LNN-based reference ($LNNr$) values, defined as
\beq
N_{f,cr,LNNr} \equiv r_{cr}N_c \ . 
\label{nfcr_lnnr}
\eeq
One does not, {\it a priori}, expect very precise agreement between
the $N_{f,cr}$ values for finite (and rather small) $N_c$ with $N_{f,cr,LNNr}$,
since these $N_c$ are far from the LNN limit $N_c \to \infty$. 
Taking the central value in Eq. (\ref{rcr_lnn}), we have 
\beq
{\rm SU}(2): \quad N_{f,cr,LNNr} = 5.8
\label{nfcr_lnnr_su2}
\eeq
\beq
{\rm SU}(3): \quad N_{f,cr,LNNr} = 8.7
\label{nfcr_lnnr_su3}
\eeq
\beq
{\rm SU}(4): \quad N_{f,cr,LNNr} = 11.6 
\label{nfcr_lnnr_su4}
\eeq
These are all in approximate agreement with the estimates of $N_{f,cr}$ in
Eqs. (\ref{nfcr_fund_su2}), (\ref{nfcr_fund_su3}), and (\ref{nfcr_fund_su4})
given above.  This shows that the approach to the LNN limit appears to be 
rather rapid, even for relatively small values of $N_c$. 

In Table \ref{kappaj_pade_lnn} we list the higher-order
$\kappa_{j,F,LNN,[0,3]}$ with $5 \le j \le 10$ calculated via a Taylor series
expansion of $\gamma_{\bar\psi\psi,F,LNN,[0,3]}$. We focus on this approximant,
since it is the only one free from poles in the disks $|\Delta_r| <
\Delta_{r,max}$ and $|\Delta_r| < \Delta_{r,cr}$.  As was the case with the
specific SU($N_c$) theories with $R=F$ that we have studied above, all of these
higher-order coefficients are positive.  We have further verified this
positivity up to a much higher order, $j=200$.


\subsection{Adjoint Representation} 

For SU($N_c$) and $R=A$, the adjoint representation, the values of $N_\ell$,
$N_u$, and $\Delta_{f,max}$ were given above in Eq. (\ref{nuell_adj}). As
indicated in Table \ref{intervals}, there is only a single integral value of
$N_f$ in the interval $I$ for these theories, namely $N_f=2$. The SU(2) theory
with $R=A$ and $N_f=2$ has been of some previous theoretical interest
\cite{sannino_su2adj,fourfermion}. As we have discussed before \cite{bvh},
since this is a real representation, one could also take $N_f$ to be
half-integral, corresponding to Majorana fermions, but the $N_f=2$ value will
be sufficient for our study here. In \cite{bvh} we carried out $n$-loop
calculations of $\gamma_{\bar\psi\psi,IR,n\ell}$ for $2 \le n \le 4$. Our two
highest-order values for SU(2) and SU(3) were
\beqs
{\rm SU}(2), \ N_f=2: \quad &&\gamma_{IR,A,3\ell}=0.543 \ , \cr\cr
                            &&\gamma_{IR,A,4\ell}=0.500 \ .
\label{gamma_su2_adjoint_Nf2_34loop}
\eeqs
and
\beqs
{\rm SU}(3), \ N_f=2: \quad &&\gamma_{IR,A,3\ell}=0.543 \ , \cr\cr
                            &&\gamma_{IR,A,4\ell}=0.523 \ .
\label{gamma_su3_adjoint_Nf2_34loop}
\eeqs

We gave explicit analytic results for the scheme-independent expansion
coefficients $\kappa_{j,A}$ with $1 \le j \le 3$ in \cite{dex} and for $j=4$ in
\cite{dexs}. Both $\kappa_{1,A}=4/9$ and $\kappa_{2,A}=341/1458$ are
independent of $N_c$, while the $k_{j,A}$ for $j \ge 3$ depend on $N_c$. 
With the prefactor $\kappa_{1,A}\Delta_f$ extracted and the Pad\'e
approximants normalized to unity at $\Delta_f=0$ as in
Eq. (\ref{gamma_pqpade}), we calculate the following approximants to 
$\gamma_{\bar\psi\psi,IR,A,\Delta_f^4}$ (in addition to
$\gamma_{\bar\psi\psi,IR,A,[3,0]} = \gamma_{\bar\psi\psi,IR,A,\Delta_f^4}$.)

For SU(2) we find
\begin{widetext} 
\beq
\gamma_{\bar\psi\psi,IR,A,[2,1]} = \frac{4}{9}\Delta_f \Bigg [ 
\frac{1 -0.793689\Delta_f -0.450643\Delta_f^2}{1-1.319924\Delta_f} \Bigg ] \ ,
\label{gamma_adjoint_su2_pade21}
\eeq
\beq
\gamma_{\bar\psi\psi,IR,A,[1,2]} = \frac{4}{9}\Delta_f \Bigg [ 
\frac{1 +6.397597\Delta_f}
     {1+5.871362\Delta_f - 3.333660\Delta_f^2} \Bigg ] \ ,
\label{gamma_adjoint_su2_pade12}
\eeq
\beq
\gamma_{\bar\psi\psi,IR,A,[0,3]} = \frac{4}{9}\Delta_f \Bigg [\frac{1}
{1-0.526235\Delta_f + 0.0329766\Delta_f^2 - 0.210971\Delta_f^3} \Bigg ] \ .
\label{gamma_adjoint_su2_pade03}
\eeq
\end{widetext}
Both the [2,1] and [0,3] Pad\'e approximants have poles in the interval $I$, so
we do not use them.  The [1,2] Pad\'e approximant has poles at
$\Delta_f=-0.1564$ and $\Delta_f=1.918$, i.e., at $N_f=2.906$ and $N_f=0.8323$,
respectively.  The first of these poles lies above $N_u$, and the second lies
below $N_\ell$, so we can use this approximant.  We list the resultant value,
$\gamma_{\bar\psi\psi,IR,A,[1,2]}=0.548$ for $N_f=2$ in Table
\ref{gamma_adjoint_values}.  Both this value from the [1,2] Pad\'e approximant
and the values $\gamma_{\bar\psi\psi,IR,A,\Delta_f^s}$ with $s=3$ and $s=4$ are
close to our previous higher-order $n$-loop calculations presented in
\cite{bvh}, given above in Eq. (\ref{gamma_su2_adjoint_Nf2_34loop}). All of
these values, which are in accord with each other, are well below the
conformality upper limit $\gamma_{\bar\psi\psi,IR} \le 2$.  

For SU(3) we calculate 
\begin{widetext}
\beq
\gamma_{\bar\psi\psi,IR,A,[2,1]} = \frac{4}{9}\Delta_f \Bigg [ 
\frac{1 -0.434579\Delta_f -0.233470\Delta_f^2}{1-0.960813\Delta_f} \Bigg ] \ ,
\label{gamma_adjoint_su3_pade21}
\eeq
\beq
\gamma_{\bar\psi\psi,IR,A,[1,2]} = \frac{4}{9}\Delta_f \Bigg [ 
\frac{1 +25.270167\Delta_f}
     {1+24.743932\Delta_f - 13.293256\Delta_f^2} \Bigg ] \ ,
\label{gamma_adjoint_su3_pade12}
\eeq
\beq
\gamma_{\bar\psi\psi,IR,A,[0,3]} = \frac{4}{9}\Delta_f \Bigg [\frac{1}
{1-0.526235\Delta_f + 0.00477966\Delta_f^2 - 0.120783\Delta_f^3} \Bigg ] \ .
\label{gamma_adjoint_su3_pade03}
\eeq
\end{widetext}
Again, both the [2,1] and [0,3] Pad\'e approximants have poles in the interval
$I$.  The [1,2] Pad\'e approximant has poles at $\Delta_f=-0.3957$ and
$\Delta_f = 1.901$, i.e., at $N_f=2.790$ and $N_f=0.8490$.  As before, the
first of these lies above $N_u$ and the second lies below $N_\ell$, so we use
this approximant.  In Table \ref{gamma_adjoint_values} we list the value of
$\gamma_{\bar\psi\psi,IR,A,[1,2]}=0.551$ for this SU(3) theory with $R=A$,
$N_f=2$.  As was the case with SU(2), both this value from the [1,2] Pad\'e
approximant and the values $\gamma_{\bar\psi\psi,IR,A,\Delta_f^s}$ with $s=3$
and $s=4$ are close to our previous higher-order $n$-loop calculations
presented in \cite{bvh} and listed above in
Eq. (\ref{gamma_su3_adjoint_Nf2_34loop}).  As was true in the SU(2) case, all
of these values are well below the
conformality upper limit $\gamma_{\bar\psi\psi,IR} \le 2$.  

From these results, we infer that the SU(2) and SU(3) theories with $R=A$ and
$N_f=2$ are in the non-Abelian Coulomb phase, and that $N_{f,cr}$ lies below
$N_f=2$ for these theories. There have been several lattice studies of the
SU(2) theory with $R=A$ and $N_f=2$, and these have also concluded that this
theory is IR-conformal \cite{lgt}.  A comparison of the values of
$\gamma_{\bar\psi\psi,IR,A}$ from these lattice measurements with our values
from our $\gamma_{\bar\psi\psi,IR,A,\Delta_f^4}$ calculation was made in
\cite{dexs,dexl}. We found that our $\gamma_{\bar\psi\psi,IR,\Delta_f^4}$ value
was in agreement with several of the lattice measurements, but noted that the
lattice measurements span a large range. These lattice results for
$\gamma_{\bar\psi\psi,IR,A}$ include the values 0.49(13) \cite{catterall},
0.22(6) \cite{deldebbio2010}, 0.31(6) \cite{degrand2011}, 0.17(5)
\cite{lsd2011}, 0.50(26) \cite{giedt2016}, and 0.20(3) \cite{tuominen2016}. 
(See these references for details of the uncertainty estimates.)  A recent work
\cite{montvay2017} reported that the value of $\gamma_{\bar\psi\psi,IR}$ that
was obtained depended on a lattice parameter (the coefficient $\beta$ of the
plaquette term in the lattice action) and hence more work was needed to
determine the actual $\gamma_{\bar\psi\psi,IR}$.


\subsection{Symmetric Rank-2 Tensor Representation} 

For our SU($N_c$) theories with $R=S_2$, the symmetric, rank-2 tensor
representation, the values of $N_\ell$, $N_u$, and $\Delta_{f,max}$ were given
above in Eq. (\ref{nuell_sym}) and are listed numerically in Table
\ref{intervals}.  In the case of SU(2), the $S_2$ representation is the same as
the adjoint representation, which we have already analyzed above.  Thus, as in 
our earlier work, we focus on the two illustrative theories, SU(3) and SU(4).
With both of these theories, the interval $I$ contains two integral values of
$N_f$, namely $N_f=2$ and $N_f=3$.  Explicit analytic results for
$\kappa_{j,S_2}$ in SU($N_c$) theories with $R=S_2$ were given for $1 \le j \le
3$ in \cite{dex} and for $j=4$ in \cite{dexl}.  The lowest-order coefficient
is %
\beq
\kappa_{1,S_2} = \frac{4(N_c-1)(N_c+2)^2}{N_c(18N_c^2+11N_c-22)} \ .
\label{kappa1_sym}
\eeq
This has the values $\kappa_{1,S_2}=200/519=0.385356$ for SU(3) and 
$\kappa_{1,S_2}=54/155=0.348387$ for SU(4). 

With the prefactor $\kappa_{1,S_2}\Delta_f$ extracted and the
Pad\'e approximant normalized as in Eq. (\ref{gamma_pqpade}), we calculate the
following Pad\'e approximants, in addition to
$\gamma_{\bar\psi\psi,IR,S_2,[3,0]} = \gamma_{\bar\psi\psi,IR,S_2,\Delta_f^4}$.
For SU(3) we obtain
\begin{widetext}
\beq
\gamma_{\bar\psi\psi,IR,S_2,[2,1]} = \frac{200}{519}\Delta_f \Bigg [ 
\frac{1 -0.327520\Delta_f -0.137718\Delta_f^2}{1-0.769654\Delta_f} \Bigg ] \, 
\eeq
\beq
\gamma_{\bar\psi\psi,IR,S_2,[1,2]} = \frac{200}{519}\Delta_f \Bigg [ 
\frac{1-8.916418\Delta_f}{1-9.358552\Delta_f + 3.935165\Delta_f^2} \Bigg ] \ , 
\label{gamma_sym_su3_pade12}
\eeq
\beq
\gamma_{\bar\psi\psi,IR,S_2,[0.3]} = \frac{200}{519}\Delta_f \Bigg [ \frac{1}
{1 -0.442134\Delta_f - 0.00708939\Delta_f^2 -0.0632120\Delta_f^2 } \Bigg ] \ . 
\label{gamma_sym_su3_pade03}
\eeq
\end{widetext} 
All of these three Pad\'e approximants with $q \ne 0$ for the SU(3) theory with
$R=S_2$ have poles in the interval $I$.  For the corresponding SU(4) theory, we
find that the Pad\'e approximants share the property with our SU(3)
results of all having poles in the interval $I$.  Consequently, for these
theories, our Pad\'e analysis does not add to our previous study of the
$\gamma_{\bar\psi\psi,IR,S_2}$ to $O(\Delta_f^3)$ in \cite{dex} and to
$O(\Delta_f^4)$ in \cite{dexs,dexl}.  There have been several lattice studies
of the SU(3) theory with $N_f=2$ fermions in the symmetric rank-2 tensor
(sextet) representation. These include Ref. \cite{degrand_sextet}, which
concluded that it is IR-conformal and obtained $\gamma_{IR} < 0.45$ and
Ref. \cite{kuti_sextet}, which concluded that it is not IR-conformal, but
instead exhibits spontaneous chiral symmetry breaking and obtained an effective
$\gamma_{IR} \simeq 1$.


\section{Pad\'e Approximants for $\beta'_{IR}$ using the $O(\Delta_f^5)$ 
Series} 
\label{betaprime_section}

\subsection{General} 

In this section we report our computation and analysis of Pad\'e approximants
for $\beta'_{IR}$, using our calculation of $\beta'_{IR}$ to $O(\Delta_f^4)$ in
\cite{dex} and to $O(\Delta_f^5)$ in \cite{dexs,dexl}. As noted, $\beta'_{IR}$
is equivalent to the anomalous dimension of ${\rm Tr}(F_{\mu\nu}F^{\mu\nu})$
\cite{traceanomaly}.  We also discuss the behavior of $\beta'_{IR}$ toward the
lower end of the non-Abelian Coulomb phase at $N_f=N_{f,cr}$. 
This behavior is relevant to the
change in the properties of the theory as $N_f$ increases through $N_{f,cr}$
\cite{pallante}. 

For a given truncation of the series (\ref{betaprime_delta_series}) to
maximal order $s$ we write $\beta'_{IR}$ as
\beqs
\beta'_{IR,\Delta_f^s} &=&
\sum_{j=2}^s d_j \Delta_f^j \cr\cr
&=&  d_2 \Delta_f^2 \Big [ 
1 + \frac{1}{d_2} \sum_{j=3}^s d_j \Delta_f^{j-2} \Big ]  \ , 
\label{betaprime_reduced}
\eeqs
and calculate the $[p,q]$ Pad\'e approximant to the expression in square
brackets, with $p+q=s-2$.  For a given fermion representation $R$, 
we denote the resultant expression using the 
$[p,q]$ Pad\'e approximant as $\beta'_{IR,[p,q]}$.  Note that the 
$[s-2,0]$ Pad\'e approximant is just the series itself, i.e., 
\beq
\beta'_{IR,[s-2,0]} = \beta'_{IR,\Delta_f^s} \ . 
\label{betaprime_jm2pade_relation}
\eeq
Thus, in particular, $\beta'_{IR,[3,0]} = \beta'_{IR,\Delta_f^5}$.  We do not
consider this, since we have already obtained evaluations of this series
truncation in previous work \cite{dexs,dexl}.

We focus here on SU($N_c$) theories with 
fermions in the fundamental, $R=F$ and consider two illustrative values of
$N_c$, namely 2 and 3.  For SU($N_c$), with $R=F$, 
\beq
d_{2,F} = \frac{16}{9(25N_c^2-11)} \ . 
\label{d2_fund}
\eeq
This takes the values $d_{2,F}=16/801 = 1.997503 \times 10^{-2}$ for SU(2) and 
$d_{2,F}=8/963=0.830737 \times 10^{-2}$ for SU(3). 

Although the lowest two
nonzero coefficients $d_2$ and $d_3$ are manifestly positive for any gauge
group $G$ and fermion representation $R$, our calculation of 
$d_4$ in \cite{dex} and $d_5$ in \cite{dexs,dexl} showed that these are both
negative for SU($N_c$) and $R=F$.  Explicitly, for SU(2) and SU(3)
 \cite{dexs,dexl}, 
\begin{widetext}
\beqs
{\rm SU}(2): \quad \beta'_{IR,F,\Delta_f^5} & =& \Delta_f^2 \Big [
  (1.99750 \times 10^{-2}
+ (3.66583 \times 10^{-3})\Delta_f
- (3.57303 \times 10^{-4})\Delta_f^2
- (2.64908 \times 10^{-5})\Delta_f^3 \ \Big ] \cr\cr
& &
\label{betaprime_sunf_p5_su2}
\eeqs
\beqs
{\rm SU}(3): \quad \beta'_{IR,F,\Delta_f^5} & =& \Delta_f^2 \Big [
  (0.83074 \times 10^{-2})
+ (0.98343 \times 10^{-3}\Delta_f
- (0.46342 \times 10^{-4})\Delta_f^2
- (0.56435 \times 10^{-5})\Delta_f^3 \ \Big ] \cr\cr
& &
\label{betaprime_sunf_p5_su3}
\eeqs
\end{widetext}
%


\subsection{SU(2)} 

For SU(2) we calculate the following $[p,q]$ Pad\'e approximants to
$\beta'_{IR,F,\Delta_f^5}$ with $q \ne 0$: 
\begin{widetext}
\beq
\beta'_{IR,F,[2,1]} = \frac{16}{801}\Delta_f^2 \Bigg [ 
\frac{1 + 0.1093795\Delta_f -0.0314939\Delta_f^2}{1-0.0741411\Delta_f} \Bigg ] 
\ , 
\label{betaprime_fund_su2_pade21}
\eeq
\beq
\beta'_{IR,F,[1,2]} = \frac{16}{801}\Delta_f^2 \Bigg [ 
\frac{1 + 0.221462\Delta_f}
{1+0.0379412\Delta_f+0.0109245\Delta_f^2} \Bigg ] \ , 
\label{betaprime_fund_su2_pade12}
\eeq
\beq
\beta'_{IR,F,[0,3]} = \frac{16}{801}\Delta_f^2 \Bigg [ \frac{1}
{1 -0.183521\Delta_f +0.0515673\Delta_f^2 -0.0114202\Delta_f^3} \Bigg ] \ .
\label{betaprime_fund_su2_pade03}
\eeq
\end{widetext} 
The [2,1] Pad\'e approximant in (\ref{betaprime_fund_su2_pade21}) has a pole at
$\Delta_f=13.488$, i.e., $N_f=-2.388$, which is not relevant. The [1,2] Pad\'e
approximant in (\ref{betaprime_fund_su2_pade12}) has a complex-conjugate pair
of poles at $\Delta_f = -1.7365 \pm 9.409i$, of magnitude $|\Delta_f|=9.5675$,
considerably larger than $\Delta_{f,max}=5.449$ and the inferred $\Delta_{f,cr}
\simeq 5 - 6$.  Hence, we can use both the [2,1] and [1,2] Pad\'e approximants
for our analysis. The [0,3] approximant in (\ref{betaprime_fund_su2_pade03})
has a real pole at $\Delta_f=4.8906$, i.e., $N_f=6.1094$, which lies in the
interval $I$ (and also has complex poles at $\Delta_f = -0.18158 \pm 4.2272i$,
with magnitude 4.231, smaller than $\Delta_{f,max}$ and $\Delta_{f,cr}$). 
Consequently, we do not use this [0,3] Pad\'e approximant.  

In Table \ref{betaprime_fund_values} we list the values of
$\beta'_{IR,F,[2,1]}$ and $\beta'_{IR,F,[1,2]}$, together with the values of
$\beta'_{IR,F,\Delta_f^j}$ with $2 \le j \le 5$, as functions of $N_f$. As is
evident, the values obtained from these Pad\'e approximants are in good
agreement with the values obtained from our series expansions in
$\Delta_f$. For example, for this SU(2) theory with $N_f=8$, 
$\beta'_{IR,F,[2,1]}=0.242$ and $\beta'_{IR,F,[1,2]}=0.247$, close to 
$\beta'_{IR,F,\Delta_f^4}=0.250$ and $\beta'_{IR,F,\Delta_f^4}=0.243$. 


\subsection{SU(3)} 

For SU(3), we calculate 
\begin{widetext}
\beq
\beta'_{IR,F,[2,1]} = \frac{8}{963}\Delta_f^2 \Bigg [ 
\frac{1 -0.00339979\Delta_f -0.0199947\Delta_f^2}{1-0.121780\Delta_f} \Bigg ] 
\ , 
\label{betaprime_fund_su3_pade21}
\eeq
\beq
\beta'_{IR,F,[1,2]} = \frac{8}{963}\Delta_f^2 \Bigg [ 
\frac{1 + 0.117412\Delta_f}{
1-(0.968004 \times 10^{-3})\Delta_f +0.00569298\Delta_f^2} \Bigg ] 
\ , 
\label{betaprime_fund_su3_pade12}
\eeq
\beq
\beta'_{IR,F,[0,3]} = \frac{8}{963}\Delta_f^2 \Bigg [ \frac{1}
{1 -0.118380\Delta_f +0.0195922\Delta_f^2 -0.00230036\Delta_f^3} \Bigg ] \ .
\label{betaprime_fund_su3_pade03}
\eeq
\end{widetext} 
The [2,1] Pad\'e approximate in (\ref{betaprime_fund_su3_pade21}) has a pole at
$\Delta_f = 8.2115$, i.e., $N_f=8.2885$, which lies in the interval $I$.  The
[1,2] approximant in (\ref{betaprime_fund_su3_pade12}) has a complex-conjugate
pair of poles at $\Delta_f = 0.0850 \pm 13.253i$, with magnitude 13.253, larger
than $\Delta_{f,max}=8.447$ and the inferred $\Delta_{f,cr} \simeq 8$.
Finally, the [0,3] approximant in (\ref{betaprime_fund_su2_pade03}) has a real
pole at $\Delta_f=8.488$, i.e., $N_f=8.012$, very close to $N_\ell=8.05$, and a
complex-conjugate pair of poles at $\Delta_f = 0.0145092 \pm 7.156461i$, with
magnitude 7.156, and hence lying within both the disk $|\Delta_f | <
\Delta_{f,max}$ and the disk $|\Delta_f| < \Delta_{f,cr}$ for this
theory. Hence, of these three Pad\'e approximants, we can only use the [1,2]
approximant for our study.  In Table \ref{betaprime_fund_values} we list the
values of $\beta'_{IR,F,[1,2]}$, together with the values of
$\beta'_{IR,F,\Delta_f^j}$ with $2 \le j \le 5$, as functions of $N_f$.  We
find that $\beta'_{IR,F,[1,2]}$ is close to the higher-order values of 
$\beta'_{IR,F,\Delta_f^j}$; for example, for this SU(3) theory with $N_f=12$,
$\beta'_{IR,F,\Delta_f^5}=0.228$, while $\beta'_{IR,F,[1,2]}=0.231$.

In \cite{dexl} we compared our scheme-independent calculations of $\beta'_{IR}$
up to $O(\Delta_f^5)$ with a lattice measurement,
$\beta'_{IR,F}=0.26(2)$ \cite{ah3} for this 
SU(3) theory with $R=F$ and $N_f=12$, finding agreement.
Here, we extend this comparison with the new
input from our Pad\'e calculation.  We recall that the conventional 
higher-order $n$-loop calculations in powers of $\alpha$ are 
$\beta'_{IR,3\ell,F}=0.2955$ and $\beta'_{IR,4\ell,F}=0.282$ \cite{bc}, which 
agree with this lattice measurement.  As we noted in \cite{dexl}, 
our higher-order scheme-independent values, namely,
$\beta'_{IR,\Delta_f^3,F}=0.258$, $\beta'_{IR,\Delta_f^4,F}=0.239$, and
$\beta'_{IR,\Delta_f^5,F}=0.228$, are also in agreement with this lattice
value from \cite{ah3}. Our new Pad\'e value, 
$\gamma_{\bar\psi\psi,IR,F,[1,2]}=0.231$, again in reasonable agreement with
both our earlier values from our higher-order $n$-loop and scheme-independent
series expansions and with the lattice value from \cite{ah3}. 


\subsection{Discussion}

The behavior of $\beta'_{IR}$ in the middle and upper part of the non-Abelian
Coulomb phase can be accurately described by the series expansion
(\ref{betaprime_delta_series}), since $\Delta_f$ approaches zero as $N_f$
approaches $N_u$ from below.  The behavior of $\beta'_{IR}$ toward the lower
end of the NACP is also of considerable interest.  As with
$\gamma_{\bar\psi\psi,IR}$, one can gain a useful perspective concerning this
behavior from known results for a vectorial, asymptotically free ${\cal N}=1$
supersymmetric theory with a gauge group $G$ and $N_f$ chiral superfields
$\Phi_j$ and $\tilde \Phi_j$ transforming according to respective
representations $R$ and $\bar R$ of $G$ \cite{nsvz}-\cite{susyreviews}. For
this supersymmetric gauge theory, the NACP occupies the range
(\ref{nacp_susy}), and $\beta'_{IR} \to 0$ at the lower end, as well as the
upper end, of the NACP \cite{grisaru97}.  In \cite{dexss} we calculated Pad\'e
approximants for $\beta'_{IR}$ from finite series expansions in $\Delta_f$ and
found consistency with the vanishing of $\beta'_{IR}$ at the lower end of the
NACP (as well as the obvious zero of $\beta'_{IR}$ at the upper end, where
$\Delta_f \to 0$).

Returning to the non-supersymmetric theories under consideration here, although
$d_2$ and $d_3$ are manifestly positive for any $G$ and $R$, we found that
$d_{4,F}$ and $d_{5,F}$ are negative for SU($N_c$) and $R=F$ \cite{dex}.
Consequently, as $N_f$ decreases below $N_u$, i.e., $\Delta_f$ increases from
zero, $\beta'_{IR}$ is initially positive, and has positive slope.  As $N_f$
decreases further, i.e., $\Delta_f$ increases further, the negative $d_{4,F}
\Delta_f^4 + d_{5,F} \Delta_f^5$ terms become progressively more important.
From Table \ref{betaprime_fund_values}, one sees that, for the range of
$N_f$ included, $\beta'_{IR,F,\Delta_f^5}$ and the resultant Pad\'e
approximants are continuing to increase with decreasing $N_f$. Evidently, 
the negative $d_{4,F}\Delta_f^4 + d_{5,F} \Delta_f^5$ terms are not
sufficiently large in magnitude to cause $\beta'_{IR}$ to vanish at the lower
end of the NACP.  Hence, one needs to calculate the scheme-independent series
expansion for $\beta'_{IR}$, Eq. (\ref{betaprime_delta_series}), to higher
order in $\Delta_f$ to see this turnover. Insofar as the behavior
of $\beta'_{IR}$ in the ${\cal N}=1$ supersymmetric gauge theory is at least a
qualitative guide to the non-supersymmetric gauge theory, then it suggests that
the exact $\beta'_{IR}$ would reach a maximum in the NACP and then would
decrease and vanish as $N_f$ decreased to the lower end of this NACP.  
We suggest that this is a plausible behavior for the non-supersymmetric theory.


\subsection{LNN Limit}

The coefficients $\hat d_{j,F}$ with $2 \le j \le 4$ were given in \cite{dex},
and $\hat d_{5,F}$ was given in \cite{dexl}. For reference, these are
\beq
\hat d_{2,F} = \frac{2^4}{3^2 \cdot 5^2} = 0.0711111 \ ,
\label{d2hat_lnn}
\eeq
\beq
\hat d_{3,F} = \frac{416}{3^3 \cdot 5^4} = 2.465185 \times 10^{-2} \ ,
\label{d3hat_lnn}
\eeq
\beq
\hat d_{4,F} = \frac{5868512}{3^5 \cdot 5^{10}}
-\frac{5632}{3^4 \cdot 5^6} \zeta_3 = -(2.876137 \times 10^{-3}) \ , 
\label{d4hat_lnn}
\eeq
and
\beqs
\hat d_{5,F} &=& -\frac{9542225632}{3^6 \cdot 5^{14}}
- \frac{1444864}{3^5 \cdot 5^9}\zeta_3
+ \frac{360448}{3^5 \cdot 5^8}\zeta_5 \cr\cr
& = & -(1.866490 \times 10^{-3}) \ .
\label{d5hat_lnn}
\eeqs

Analogously to Eq. (\ref{betaprime_reduced}), we write
\beqs
\beta'_{IR,\Delta_r^s} &=&
\sum_{j=2}^s \hat d_{j,F} \Delta_r^j \cr\cr
&=&  \hat d_{2,F} \Delta_r^2 \Big [ 
1 + \frac{1}{\hat d_{2,F}} \sum_{j=3}^s \hat d_{j,F} \Delta_r^{j-2} \Big ] \ ,
\cr\cr
&& 
\label{betaprime_reduced_lnn}
\eeqs
and calculate the $[p,q]$ Pad\'e approximant to the expression in square
brackets, with $p+q=s-2$.  We have calculated analytic results for Pad\'e
approximants to $\gamma_{\bar\psi\psi,IR,F,\Delta_r^5}$. As before, we 
present these in numerical form. We find 
\begin{widetext}
\beq
\beta'_{IR,F,LNN,[2,1]} = \frac{16}{225}\Delta_r^2 \Bigg [ 
\frac{1 -0.302290\Delta_r -0.265417\Delta_r^2}{1-0.648957\Delta_r} \Bigg ] 
\ , 
\label{betaprime_fund_lnn_pade21}
\eeq
\beq
\beta'_{IR,F,LNN,[1,2]} = \frac{16}{225}\Delta_r^2 \Bigg [ 
\frac{1 + 0.270549\Delta_r}{
1-(0.0761180)\Delta_r +0.0668333\Delta_r^2} \Bigg ] 
\ , 
\label{betaprime_fund_lnn_pade12}
\eeq
\beq
\beta'_{IR,F,LNN,[0,3]} = \frac{16}{225}\Delta_r^2 \Bigg [ \frac{1}
{1 -0.346667\Delta_r +0.160623\Delta_r^2 -0.0434565\Delta_r^3} \Bigg ] \ .
\label{betaprime_fund_lnn_pade03}
\eeq
\end{widetext} 
The [2,1] Pad\'e approximant has a pole at $\Delta_r=1.54093$, i.e.,
$r=3.95907$.  This is in the interval $I$ and the inferred NACP, so we cannot
use this approximant for our work. The [1,2] Pad\'e approximant has a
complex-conjugate pair of poles at $\Delta_r = 0.56946 \pm 3.82601i$ with
magnitude $|\Delta_r|=3.86815$, which is greater than both
$\Delta_{r,max}=2.8846$ and $\Delta_{cr}=2.6$.  Hence, we can use this
approximant.  Finally, the [0,3] Pad\'e approximant has a pole at
$\Delta_r=3.36019$, i.e., $r=2.13981$ and a complex-conjugate pair of poles
at $\Delta_r=0.168003 \pm 2.611526i$ with magnitude 2.6169.  Although the real
pole is outside of the interval $I$ and the inferred NACP, the complex poles
lie within the disks $|\Delta_r| < \Delta_{r,max}$ and $|\Delta_r| <
\Delta_{cr}$, so we do not use this [0,3] approximant. In 
Table \ref{betaprime_lnn_values} we list the values of
$\beta'_{IR,F,LNN,[1,2]}$ as a function of $r$, together with our previously
calculated $\beta'_{IR,F,\Delta_r^s}$ with $2 \le s \le 5$ for comparison. We
find that the values of the [1,2] Pad\'e approximant are close to those of 
the high-order $\beta'_{IR,F,\Delta_r^s}$; for example, at $r=4.0$, 
$\beta'_{IR,F,\Delta_r^5}=0.214$ while $\beta'_{IR,F,LNN,[1,2]}=0.217$. As $r$
decreases toward the lower end of the non-Abelian Coulomb phase, 
$\beta'_{IR,F,LNN,[1,2]}$ becomes slightly larger than 
$\beta'_{IR,F,\Delta_r^5}$, just as was the case as $N_f$ decreased toward the
respective lower ends of the NACP in the specific SU(2) and SU(3) theories
discussed above. 

As with $\gamma_{\bar\psi\psi,IR,F,LNN}$, it is of interest to carry out a
Taylor series expansion of $\beta'_{IR,F,LNN,[1,2]}$ to determine its
prediction for the coefficients $\hat d_{j,F}$ with $j \ge 6$.  We have done
this and present the resultant $\hat d_{j,F}$ coefficients in Table
\ref{dj_pade_lnn}.


\section{Conclusions}
\label{conclusion_section}

In this paper, we have presented several new results on the anomalous
dimension, $\gamma_{\bar\psi\psi,IR}$, and the derivative of the beta function,
$\beta'_{IR}$, at an infrared fixed point of the renormalization group in
vectorial, asymptotically free SU($N_c$) gauge theories with $N_f$ fermions
transforming according to several representations $R$, including the
fundamental, adjoint, and rank-2 symmetric tensor.  We have used our series for
$\gamma_{\bar\psi\psi,IR}$ to $O(\Delta_f^4)$ to calculate Pad\'e approximants
and have evaluated these to obtain further estimates of
$\gamma_{\bar\psi\psi,IR}$.  Our new results using these Pad\'e approximants
are consistent with our earlier results using the series themselves calculated
to $O(\Delta_f^4)$.  We have compared the values of $\gamma_{\bar\psi\psi,IR}$
with lattice measurements for various theories.  Taylor-series expansions of
the Pad\'e approximants have been calculated to determine their predictions for
higher-order coefficients. We have found that all of the Pad\'e approximants
that we have calculated that satisfy the requisite constraints (absence of
poles in the disks $|\Delta_f| < \Delta_{f,max}$ and $|\Delta_f| <
\Delta_{f,cr}$ in the complex $\Delta_f$ plane) yield Taylor-series expansions
with positive coefficients $\kappa_j$, providing further support for our
earlier conjecture that the $\kappa_j$ are positive.  We
have also used our Pad\'e results to obtain new estimates of the value of
$N_{f,cr}$ at the lower end of the non-Abelian Coulomb phase for various $N_c$
and $R$.  Since, for a given SU($N_c$) gauge group and fermion representation
$R$, the upper end of the NACP, namely $N_u$, is known exactly, these estimates
of $N_{f,cr}$ are equivalently estimates of the extent of the non-Abelian
Coulomb phase, as a function of $N_f$, for each of the theories that we have
considered.  In a different but related application, our values of $N_{f,cr}$
are useful for the phenomenological program of constructing and studying
quasi-conformal gauge theories to explore ideas for possible ultraviolet
completions of the Standard Model. This is because, for a given gauge group $G$
and fermion representation $R$, one must choose $N_f$ to be slightly below
$N_{f,cr}$ (requiring that one know $N_{f,cr}$) in order to achieve the
quasi-conformal behavior whose spontaneous breaking via formation of fermion
condensates could have the potential to yield a light, dilatonic Higgs-like
scalar. We have carried out calculations of Pad\'e approximants for
$\beta'_{IR}$, using our series to $O(\Delta_f^5)$ for these theories. Again,
the results for $\beta'_{IR}$ obtained from these Pad\'e approximants are
consistent with, and extend, our earlier analyses using the series
themselves. Our values for $\gamma_{\bar\psi\psi,IR}$ and $\beta'_{IR}$
obtained with Pad\'e approximants provide further information about fundamental
properties of conformal field theories. Finally, we have presented new analytic
and numerical results assessing the accuracy of a series expansion of
$\gamma_{\bar\psi\psi,IR}$ to finite order in powers of $\Delta_f$ by
comparison with the exactly known expression in an ${\cal N}=1$ supersymmetric
gauge theory, showing that an expansion to $O(\Delta_f^4)$ is quite accurate
throughout the entire non-Abelian Coulomb phase of this supersymmetric theory.


\begin{acknowledgments}

The research of T.A.R. and R.S. was supported in part by the Danish National
Research Foundation grant DNRF90 to CP$^3$-Origins at SDU and by the
U.S. National Science Foundation Grant NSF-PHY-16-1620628, respectively.  

\end{acknowledgments}



\newpage

\begin{widetext}

\begin{table}
  \caption{\footnotesize{$N_\ell$, $N_u$, $\Delta_{f,max}$, and interval 
$I$ in terms of $N_f$, for
$G={\rm SU}(N_c)$ with fermions in the representation $R$ equal to fundamental
    (F), adjoint (A), and rank-2 symmetric (S$_2$) tensor.  The interval $I$ is
    listed for $N_f$ formally generalized to real numbers, 
${\mathbb R}_+$ and for physical, integral values of $N_f \in {\mathbb N}_+$.
Note that for $R=A$, $N_\ell$ and $N_u$ are independent of $N_c$.}}
\begin{center}
\begin{tabular}{|c|c|c|c|c|c|c|} \hline\hline
$N_c$ & $R$ & $N_\ell$ & $N_u$ & $\Delta_{f,max}$ & 
$I$, $N_f \in {\mathbb R}_+$ & $I$, $N_f \in {\mathbb N}_+$ \\
\hline
2    & F & 5.551 & 11 & 5.449 & $5.551 < N_f < 11$ & $6 \le N_f \le 10$ \\
\hline
3    & F & 8.053 & 16.5 & 8.447 & $8.053 < N_f < 16.5$ & $9 \le N_f \le 16$ \\
\hline
4    & F & 10.615 & 22 & 11.385 & $10.615 < N_f < 22$ & $11 \le N_f \le 21$ \\
\hline\hline
$N_c$& A & 1.0625 & 2.75& 1.6875 & $1.0625 < N_f < 1.6875$ & $N_f=2$ \\
\hline\hline
3 & S$_2$ & 1.224 & 3.3000 & 2.076 & $1.224 < N_f < 3.300$ & $N_f=2, \ 3$ \\
4 & S$_2$ & 1.353 & 3.3667 & 2.313 & $1.353 < N_f < 3.667$ & $N_f=2, \ 3$ \\
\hline\hline
\end{tabular}
\end{center}
\label{intervals}
\end{table}


\begin{table}
\caption{\footnotesize{Values of $\gamma_{\bar\psi\psi,IR,F,[1,2]}$ and 
$\gamma_{\bar\psi\psi,IR,F,[0,3]}$, as functions of $N_f$ for 
$G={\rm SU}(N_c)$ with $2 \le N_c \le 4$ and fermion representation 
and $R=F$. For comparison, we include values of 
$\gamma_{\bar\psi\psi,IR,F,\Delta_f^s}$ with $1 \le s \le 4$. 
Values of anomalous dimensions that exceed the conformality upper
bound $\gamma_{\bar\psi\psi,IR} \le 2$ are marked with brackets.
To save space, we omit the $\bar\psi\psi$ from the subscripts in the table,
writing $\gamma_{\bar\psi\psi,IR,F,\Delta_f^s} \equiv 
\gamma_{IR,F,\Delta_f^s}$, etc.}}
\begin{center}
\begin{tabular}{|c|c|c|c|c|c|c|c|} \hline\hline
$N_c$ & $N_f$ & $\gamma_{IR,F,\Delta_f}$ & $\gamma_{IR,F,\Delta_f^2}$ 
& $\gamma_{IR,F,\Delta_f^3}$ & $\gamma_{IR,F,\Delta_f^4}$ 
& $\gamma_{IR,F,[1,2]}$ & $\gamma_{IR,F,[0,3]}$ 
\\ \hline
2 & 6  & 0.337  & 0.520  & 0.596  & 0.698  & 0.711 & 1.090    \\
2 & 7  & 0.270  & 0.387  & 0.426  & 0.467  & 0.466 & 0.541    \\
2 & 8  & 0.202  & 0.268  & 0.285  & 0.298  & 0.296 & 0.310    \\
2 & 9  & 0.135  & 0.164  & 0.169  & 0.172  & 0.171 & 0.173    \\
2 & 10 & 0.0674 & 0.0747 & 0.07535& 0.0755 &0.0755 & 0.0755   \\
\hline
3 & 8  & 0.424  & 0.698  & 0.844  & 1.036  & 1.149  &[2.848]  \\
3 & 9  & 0.374  & 0.587  & 0.687  & 0.804  & 0.844  & 1.2645  \\
3 & 10 & 0.324  & 0.484  & 0.549  & 0.615  & 0.627  & 0.764   \\
3 & 11 & 0.274  & 0.389  & 0.428  & 0.462  & 0.464  & 0.5105  \\
3 & 12 & 0.224  & 0.301  & 0.323  & 0.338  & 0.3375 & 0.352   \\
3 & 13 & 0.174  & 0.221  & 0.231  & 0.237  & 0.236  & 0.240   \\
3 & 14 & 0.125  & 0.148  & 0.152  & 0.153  & 0.153  & 0.154   \\
3 & 15 & 0.0748 & 0.0833 & 0.0841 & 0.0843 & 0.0843 & 0.0843  \\
3 & 16 & 0.0249 & 0.0259 & 0.0259 & 0.0259 & 0.0259 & 0.0259  \\
\hline
4 &11  & 0.424  & 0.694  & 0.844 & 1.029   & 1.138  &[2.491]  \\
4 &12  & 0.386  & 0.609  & 0.721 & 0.8475  & 0.897  & 1.376   \\
4 &13  & 0.347  & 0.528  & 0.610 & 0.693   & 0.713  & 0.914   \\
4 &14  & 0.308  & 0.451  & 0.509 & 0.561   & 0.568  & 0.656   \\
4 &15  & 0.270  & 0.379  & 0.418 & 0.448   & 0.450  & 0.488   \\
4 &16  & 0.231  & 0.312  & 0.336 & 0.352   & 0.352  & 0.368   \\
4 &17  & 0.193  & 0.249  & 0.263 & 0.2705  & 0.270  & 0.276   \\
4 &18  & 0.154  & 0.190  & 0.197 & 0.200   & 0.200  & 0.202   \\
4 &19  & 0.116  & 0.136  & 0.139 & 0.140   & 0.140  & 0.140   \\
4 &20  & 0.0771 & 0.0860 & 0.0869 & 0.0871 & 0.0871 & 0.0872  \\
4 &21  & 0.0386 & 0.0408 & 0.0409& 0.0409  & 0.0409 & 0.0409  \\
\hline\hline
\end{tabular}
\end{center}
\label{gamma_fund_values}
\end{table}
%


\begin{table}
  \caption{\footnotesize{Values of $\gamma_{\bar\psi\psi,IR,F,[p,q]}$ and, 
for comparison, values of $\gamma_{\bar\psi\psi,IR,F,\Delta_r^s}$ with 
$1 \le s \le 4$, in the LNN limit (\ref{lnn}). 
Values of anomalous dimensions that exceed the conformality upper
bound $\gamma_{\bar\psi\psi,IR} \le 2$ are marked with brackets.
To save space, we omit the $\bar\psi\psi$ from the subscripts in the table,
writing $\gamma_{\bar\psi\psi,IR,F,\Delta_r^s} \equiv 
\gamma_{IR,F,\Delta_r^s}$, etc.}}
\begin{center}
\begin{tabular}{|c|c|c|c|c|c|c|} \hline\hline
$r$ 
& $\gamma_{IR,F,\Delta_r}$
& $\gamma_{IR,F,\Delta_r^2}$
& $\gamma_{IR,F,\Delta_r^3}$
& $\gamma_{IR,F,\Delta_r^4}$
& $\gamma_{IR,F,[1,2]}$
& $\gamma_{IR,F,[0,3]}$
\\ \hline
2.8& 0.432 & 0.706 & 0.870  & 1.064  & 1.176  & [2.608]  \\
2.9& 0.416 & 0.670 & 0.817  & 0.983  & 1.065  & 1.968    \\
3.0& 0.400 & 0.635 & 0.765  & 0.908  & 0.966  & 1.567    \\
3.2& 0.368 & 0.567 & 0.668  & 0.770  & 0.798  & 1.087    \\
3.4& 0.336 & 0.502 & 0.579  & 0.650  & 0.662  & 0.809    \\
3.6& 0.304 & 0.440 & 0.497  & 0.5445 & 0.548  & 0.624    \\
3.8& 0.272 & 0.381 & 0.422  & 0.452  & 0.452  & 0.491    \\
4.0& 0.240 & 0.325 & 0.353  & 0.371  & 0.370  & 0.389    \\
4.2& 0.208 & 0.272 & 0.290  & 0.300  & 0.299  & 0.308    \\
4.4& 0.176 & 0.2215& 0.233  & 0.238  & 0.237  & 0.241    \\
4.6& 0.144&0.1745& 0.1805   & 0.183  & 0.1825 & 0.184    \\
4.8& 0.112 & 0.130 & 0.133  & 0.134  & 0.134  & 0.134    \\
5.0& 0.0800& 0.0894& 0.09045& 0.0907 & 0.0906 & 0.0907   \\
5.2& 0.0480& 0.0514& 0.0516 & 0.0516 & 0.0516  & 0.0516  \\
5.4& 0.0160& 0.0164& 0.0164 & 0.0164 & 0.0164  & 0.0164  \\
\hline\hline
\end{tabular}
\end{center}
\label{gamma_lnn_values}
\end{table}
%


\begin{table}
  \caption{\footnotesize{Values of higher-order $\kappa_{j,F,[0,3]}$ with
$5 \le j \le 10$ 
  obtained by Taylor series expansions of $\gamma_{\bar\psi\psi,F,[0,3]}$ for
  SU($N_c$) theories with $N_c=2, \ 3, \ 4$. 
 The notation $a$e-n means $a \times 10^{-n}$.}}
\begin{center}
\begin{tabular}{|c|c|c|c|} \hline\hline
$j$ &
$\kappa_{j,F,[0,3],{\rm SU}(2)}$ & 
$\kappa_{j,F,[0,3],{\rm SU}(3)}$ &
$\kappa_{j,F,[0,3],{\rm SU}(4)}$ 
\\ \hline
5  & 2.876447e-5  & 0.427629e-5    & 1.095416e-6  \\
6  & 3.721786e-6  & 0.395378e-6    & 0.7862105e-7 \\
7  & 0.606848e-6  & 0.423869e-7    & 0.640937e-8  \\
8  & 1.056182e-7  & 0.475954e-8    & 0.539017e-9  \\
9  & 1.625598e-8  & 0.498035e-9    & 0.429237e-10 \\
10 & 2.5266295e-9 & 0.522418e-10   & 0.343311e-11 \\
\hline\hline
\end{tabular}
\end{center}
\label{kappaj_pade_fund}
\end{table}
%


\begin{table}
  \caption{\footnotesize{Values of higher-order $\hat\kappa_{j,F,LNN,[0,3]}$ 
  obtained by Taylor series expansions of $\gamma_{\bar\psi\psi,F,LNN,[0,3]}$ 
  in the LNN limit. The notation $a$e-n means $a \times 10^{-n}$.}}
\begin{center}
\begin{tabular}{|c|c|} \hline\hline
$j$ & $\hat\kappa_{j,F,LNN,[0,3]}$ 
\\ \hline
5  & 1.257923e-3   \\
6  & 0.378242e-3   \\
7  & 1.262231e-4   \\
8  & 0.427164e-4   \\
9  & 1.392391e-5   \\
10 & 0.456625e-5   \\
\hline\hline
\end{tabular}
\end{center}
\label{kappaj_pade_lnn}
\end{table}
%


\begin{table}
  \caption{\footnotesize{Values of $\gamma_{\bar\psi\psi,A,IR,[1,2]}$ and, 
for comparison, $\gamma_{\bar\psi\psi,A,IR,\Delta_f^s}$ with $1 \le s \le 4$,
for $N_c=2, \ 3$, $R=A$ (adjoint), and $N_f=2$. 
To save space, we omit the $\bar\psi\psi$ from the subscripts in the table,
writing $\gamma_{\bar\psi\psi,IR,A,\Delta_f^s} \equiv 
\gamma_{IR,A,\Delta_f^s}$, etc.}}
\begin{center}
\begin{tabular}{|c|c|c|c|c|c|c|} \hline\hline
$N_c$ & $N_f$ & $\gamma_{IR,A,\Delta_f}$ & $\gamma_{IR,A,\Delta_f^2}$
& $\gamma_{IR,A,\Delta_f^3}$ & $\gamma_{IR,A,\Delta_f^4}$ 
& $\gamma_{IR,[1,2]}$ 
\\ \hline
2 & 2 & 0.333 & 0.465 & 0.511 & 0.556 & 0.548  \\
3 & 2 & 0.333 & 0.465 & 0.516 & 0.553 & 0.551  \\
\hline\hline
\end{tabular}
\end{center}
\label{gamma_adjoint_values}
\end{table}
%


\begin{table}
  \caption{\footnotesize{Values of $\beta'_{IR,F,[2,1]}$ and 
$\beta'_{IR,F,[1,2]}$ for SU($N_c$) with $N_c=2, \ 3$ and 
$R=F$ (fundamental) as a function of $N_f$.  For comparison, we also include
the values of $\beta'_{IR,F,\Delta_f^s}$ from \cite{dex,dexl}. 
The  columns list $\beta'_{IR,F,\Delta_f^s}$ for $2 \le j \le 5$,
and then $\beta'_{IR,F,[2,1]}$ and $\beta'_{IR,F,[1,2]}$. As discussed in 
the text, for SU(3), we use only the [1,2] Pad\'e approximant.}}
\begin{center}
\begin{tabular}{|c|c|c|c|c|c|c|c|c|} \hline\hline
$N_c$ & $N_f$ & $\beta'_{IR,F,\Delta_f^2}$ & $\beta'_{IR,F,\Delta_f^3}$ 
& $\beta'_{IR,F,\Delta_f^4}$ & $\beta'_{IR,F,\Delta_f^5}$
& $\beta'_{IR,F,[2,1]}$ & $\beta'_{IR,F,[1,2]}$
\\ \hline
2 &  6 & 0.499 & 0.957 & 0.734 & 0.6515 & 0.603  & 0.719  \\
2 &  7 & 0.320 & 0.554 & 0.463 & 0.436  & 0.424  & 0.454  \\
2 &  8 & 0.180 & 0.279 & 0.250 & 0.243  & 0.242  & 0.247  \\
2 &  9 & 0.0799& 0.109 & 0.1035& 0.103  & 0.1025 & 0.103  \\
2 & 10 & 0.0200& 0.0236& 0.0233& 0.0233 & 0.0233 & 0.0233 \\
\hline
3 &  9 & 0.467 & 0.882 & 0.7355& 0.602  & $-$    & 0.669   \\
3 & 10 & 0.351 & 0.621 & 0.538 & 0.473  & $-$    & 0.501   \\
3 & 11 & 0.251 & 0.415 & 0.3725& 0.344  & $-$    & 0.354   \\
3 & 12 & 0.168 & 0.258 & 0.239 & 0.228  & $-$    & 0.231   \\
3 & 13 & 0.102 & 0.144 & 0.137 & 0.134  & $-$    & 0.135   \\
3 & 14 & 0.0519& 0.0673& 0.0655& 0.0649 & $-$    & 0.0650  \\
3 & 15 & 0.0187& 0.0220& 0.0218& 0.0217 & $-$    & 0.0217   \\
3 & 16 & 2.08e-3&2.20e-3&2.20e-3&2.20e-3& $-$    & 2.20e-3  \\
\hline\hline
\end{tabular}
\end{center}
\label{betaprime_fund_values}
\end{table}
%

%

\begin{table}
  \caption{\footnotesize{Values of $\beta'_{IR,F,LNN,[1,2]}$ 
in the LNN limit, as a function of the ratio $r$
defined in Eq. (\ref{lnn}). For comparison, we also include
the values of $\beta'_{IR,F,LNN,\Delta_r^s}$ from \cite{dex,dexl} for 
$2 \le s \le 5$.}}
\begin{center}
\begin{tabular}{|c|c|c|c|c|c|} \hline\hline
$r$ & $\beta'_{IR,F,LNN,\Delta_r^2}$ & $\beta'_{IR,F,LNN,\Delta_r^3}$ 
& $\beta'_{IR,F,LNN,\Delta_r^4}$ & $\beta'_{IR,F,LNN,\Delta_r^5}$
& $\beta'_{IR,F,LNN,[1,2]}$ 
\\ \hline
2.8& 0.518 & 1.004 & 0.851  & 0.583  & 0.700   \\
3.0& 0.444 & 0.830 & 0.717  & 0.535  & 0.607   \\
3.2& 0.376 & 0.676 & 0.596  & 0.475  & 0.518   \\
3.4& 0.314 & 0.542 & 0.486  & 0.410  & 0.433   \\
3.6& 0.257 & 0.426 & 0.388  & 0.342  & 0.354   \\
3.8& 0.2055& 0.327 & 0.303  & 0.276  & 0.282   \\
4.0& 0.160 & 0.243 & 0.229  & 0.214  & 0.217   \\
4.2& 0.120 & 0.174 & 0.166  & 0.159  & 0.160   \\
4.4& 0.0860& 0.119 & 0.115  & 0.112  & 0.112   \\
4.6& 0.0576& 0.0756& 0.0737 & 0.0726 & 0.0727  \\
4.8& 0.0348& 0.0433& 0.0426 & 0.0423 & 0.0423  \\
5.0& 0.0178& 0.0209& 0.0207 & 0.0206 & 0.0206  \\
5.2&0.640e-2& 0.707e-2&0.704e-2&0.704e-2 & 0.704e-2 \\
5.4&0.711e-3&0.736e-3&0.735e-3&0.735e-3  & 0.735e-3 \\
\hline\hline
\end{tabular}
\end{center}
\label{betaprime_lnn_values}
\end{table}
%

\begin{table}
  \caption{\footnotesize{Values of higher-order $\hat d_{j,F,LNN,[1,2]}$ 
  obtained by Taylor series expansions of $\beta'_{IR,F,LNN,[1,2]}$ 
  in the LNN limit. The notation $a$e-n means $a \times 10^{-n}$.}}
\begin{center}
\begin{tabular}{|c|c|} \hline\hline
$j$ & $\hat d_{j,F,LNN,[1,2]}$ 
\\ \hline
6  & 0.5014815e-4     \\
7  & 1.285608e-4      \\
8  & 0.643423e-5       \\
9  & $-0.8102375$1e-5 \\
10 & $-1.046757$e-6   \\
11 & 0.461831e-6      \\
12 & 1.051119e-7      \\
\hline\hline
\end{tabular}
\end{center}
\label{dj_pade_lnn}
\end{table}

\end{widetext}


\begin{thebibliography}{99}
\bibitem{fm}
%
The assumption of massless fermions incurs no loss of generality,
since if a fermion had a nonzero mass $m_0$, it would be integrated out of
the effective field theory at scales $\mu < m_0$, and hence would not affect
the IR limit $\mu \to 0$ considered here.
\bibitem{nfintegral}
%
Here and elsewhere, when an expression is given for $N_f$
that formally evaluates to a non-integral real value, it is understood
implicitly that one infers an appropriate integral value from it.
\bibitem{casimir}
%
Let $T^a_R$ denote the generators of the Lie algebra of a group $G$ in the
representation $R$, where $a$ is a group index, and let $d_R$ denote the
dimension of $R$. 
The Casimir invariants $C_2(R)$ and $T_R$ are defined as follows:
$T^a_RT^a_R = C_2(R) I$, where here $I$ is 
the $d_R \times d_R$ identity matrix, and 
${\rm Tr}_R(T^a_R T^b_R) = T(R)\delta_{ab}$.  For a fermion $f$
transforming according to a representation $R$, we often use the equivalent
compact notation $T_f \equiv T(R)$ and $C_f \equiv C_2(R)$. We also use the
notation $C_A \equiv C_2(A) \equiv C_2(G)$.
\bibitem{other}
%
  In some early work there was also an attempt to apply the $\Delta_f$
  expansion to actual quantum chromodynamics (QCD); however, the properties of
  an asymptotically free gauge theory are qualitatively different
  for $\Delta_f < \Delta_{f,cr}$ in the (chirally symmetric, deconfined) 
  non-Abelian Coulomb  phase and for $\Delta_f > \Delta_{f,cr}$ in QCD,
  with its confinement and spontaneous chiral symmetry breaking.
\bibitem{schisb} 
%
  For a given $G$ and $R$, if $N_f < N_{f,cr}$, so that there is spontaneous
  chiral symmetry breaking, then even if the perturbative beta function with
  the original $N_f$ fermions exhibits an IR zero at some value $\alpha_{IR}$,
  this is only approximate, since the theory flows away from this value as the
  fermions gain dynamical mass and are integrated out, leaving a pure gluonic
  low-energy effective field theory with a different beta function.
\bibitem{nonsat}
%
  We note that, although $\gamma_{M,IR}$ saturates its conformality upper bound
  at the lower end of the NACP in this ${\cal N}=1$ supersymmetric gauge
  theory, anomalous dimensions of various other composite chiral superfield
  operator products were calculated in \cite{dexss}, and it was found that they
  do not saturate their conformality upper bounds.
\end{thebibliography}
\end{document}